\documentclass[aps,prc,twocolumn,superscriptaddress]{revtex4-2}
\usepackage{lineno,isotope,mathrsfs}
\modulolinenumbers[5]
\usepackage{amsmath,bbold,amssymb,epsfig,bm,mathrsfs,feynmp}
\usepackage{color,slashed,exscale,multirow,soul,nicefrac,gensymb}
\usepackage{longtable,times,txfonts,xcolor,graphicx,tikz}
\usepackage[normalem]{ulem}
\usepackage[breaklinks,colorlinks,citecolor=blue]{hyperref}
\definecolor{red}{rgb}{0.8,0,0}
\definecolor{violet}{rgb}{0.4,0,0.4}
\definecolor{green}{rgb}{0,0.5,0.0}
\definecolor{navy}{rgb}{0.0,0.0,0.6}
\definecolor{orange}{rgb}{0.8,0.2,0.0}

\usepackage[normalem]{ulem}  

\newcommand{\bea}{\begin{eqnarray}}
\newcommand{\eea}{\end{eqnarray}}

\begin{document}
\title{
Bayesian inferences on covariant density functionals 
from multimessenger astrophysical data: Nucleonic models
}
%
\author{Jia-Jie Li}
\email{jiajieli@swu.edu.cn}           
\affiliation{School of Physical Science and Technology, 
             Southwest University, Chongqing 400715, China}           
\author{Yu Tian}           
\affiliation{School of Physical Science and Technology, 
             Southwest University, Chongqing 400715, China}          
%
\author{Armen Sedrakian}   
\email{sedrakian@fias.uni-frankfurt.de}         
\affiliation{Frankfurt Institute for Advanced Studies,
             D-60438 Frankfurt am Main, Germany}         
\affiliation{Institute of Theoretical Physics,
            University of Wroc\l{}aw, 50-204 Wroc\l{}aw, Poland} 
\begin{abstract}
\begin{description}
\item[Background] Bayesian inference frameworks incorporating
  multi-messenger astrophysical constraints have recently been applied
  to covariant density functional (CDF) models to constrain their
  parameters. Among these, frameworks utilizing  CDFs with
  density-dependent meson-nucleon couplings furnishing the equation of
  state (EoS) of compact star (CS) matter have been explored. This class
  of CDF models has been extensively tested against various nuclear
  phenomena, providing a highly predictive description of CSs in terms 
  of nuclear characteristics defined in the vicinity of nuclear saturation 
  density and isopin symmetric limit. 
\item[Purpose] The aforementioned inference framework has not yet
  incorporated astrophysical objects with potentially extreme high
  masses or ultra-small radii among its constraints, leaving its
  flexibility and predictive power under such extreme parameters still
  unknown. We aim at exploring the capabilities of this framework to
  account for these additional data.
\item[Method] We apply the Bayesian inference framework based on CDFs
  with density dependent couplings. The astrophysical data are expanded
  to include not only the latest multi-messenger constraints from
  NICER telescope and gravitational wave events but also the highest measured
  mass to date for the ``black widow" pulsar PSR J0952-0607 and the
  mass-radius estimates for the ultra-compact, low-mass object HESS
  J1731-347.  The significance of the individual impact of these two
  sources, as well as the broader picture that emerges from their
  combined influence, is examined by analyzing the posterior
  distributions of key CS properties, nuclear characteristics at 
  saturation density, and their correlations.
\item[Results] Our systematic Bayesian analysis, which incorporates
  the PSR J0952-0607 and/or HESS J1731-347 data alongside typical
  astrophysical constraints, indicates that our CDF models can support
  higher maximum masses for CSs, reaching up to $2.4$-$2.5\,M_{\odot}$. 
  However, achieving sufficient softening of the EoS in the low-density 
  regime to accommodate the HESS J1731-347 data remains challenging. 
  Nonetheless, we are able to impose tighter constraints on the 
  parameter space of CDF models, ensuring consistency with current 
  nuclear experimental and astrophysical data.
\item[Conclusions] CDF models with density-dependent meson-nucleon
  couplings encompass a wide range of nuclear and astrophysical
  phenomena, providing a robust theoretical framework for interpreting
  compact objects. However, the predicted lower limit for the radii of
  low-mass stars is approximately 12 km, which stems from the
  restricted degrees of freedom in the isovector sector. These
  limitations can be addressed by extending CDF models to introduce
  more complex density dependence of the symmetry energy. Additionally, 
  smaller radii may be achieved by allowing for a low-density 
  deconfinement phase transition to quark matter.
\end{description}
\end{abstract}
\date{\today}
\maketitle
%
\section{Introduction}
\label{sec:Intro}
The systematic construction of dense matter equation of state (EoS) models 
and statistical inference of EoS parameters from data is an endeavor that 
has recently attracted  significant attention. The Bayesian framework has 
been applied to models covering the range from fully physics agnostic 
non-parametric models~\citep{Raaijmakers:2019,Landry:2020,Raaijmakers:2020,Legred:2021,Pang:2021,Altiparmak:2022,Annala:2022,Annala:2023,Chimanski:2022,Rutherford:2024,Fan:2024}
to microscopic models based on nuclear
potentials~\citep{Mondal:2023,Zhou:2023,Beznogov:2024a,Beznogov:2024b,Tsang:2023} 
to density functional method-based
models~\citep{Traversi:2020,Malik:2022a,Malik:2022b,Sun:2022,Zhu:2023,Providencia:2023,Char:2023,Beznogov:2023,Malik:2023,Salinas:2023,Huang:2024,Parmar:2024,Scurto:2024,Tewari:2024,Lijj:2024c,Char:2025}.
This effort has matured in the current era of multi-messenger astronomy 
with a large push to explore model-dependence in statistical inferences. The future thus 
lies in combining more and more sets of data of both types to understand 
nuclear and neutron star models better.

\begin{figure}[b]
\centering
\includegraphics[width = 0.46\textwidth]{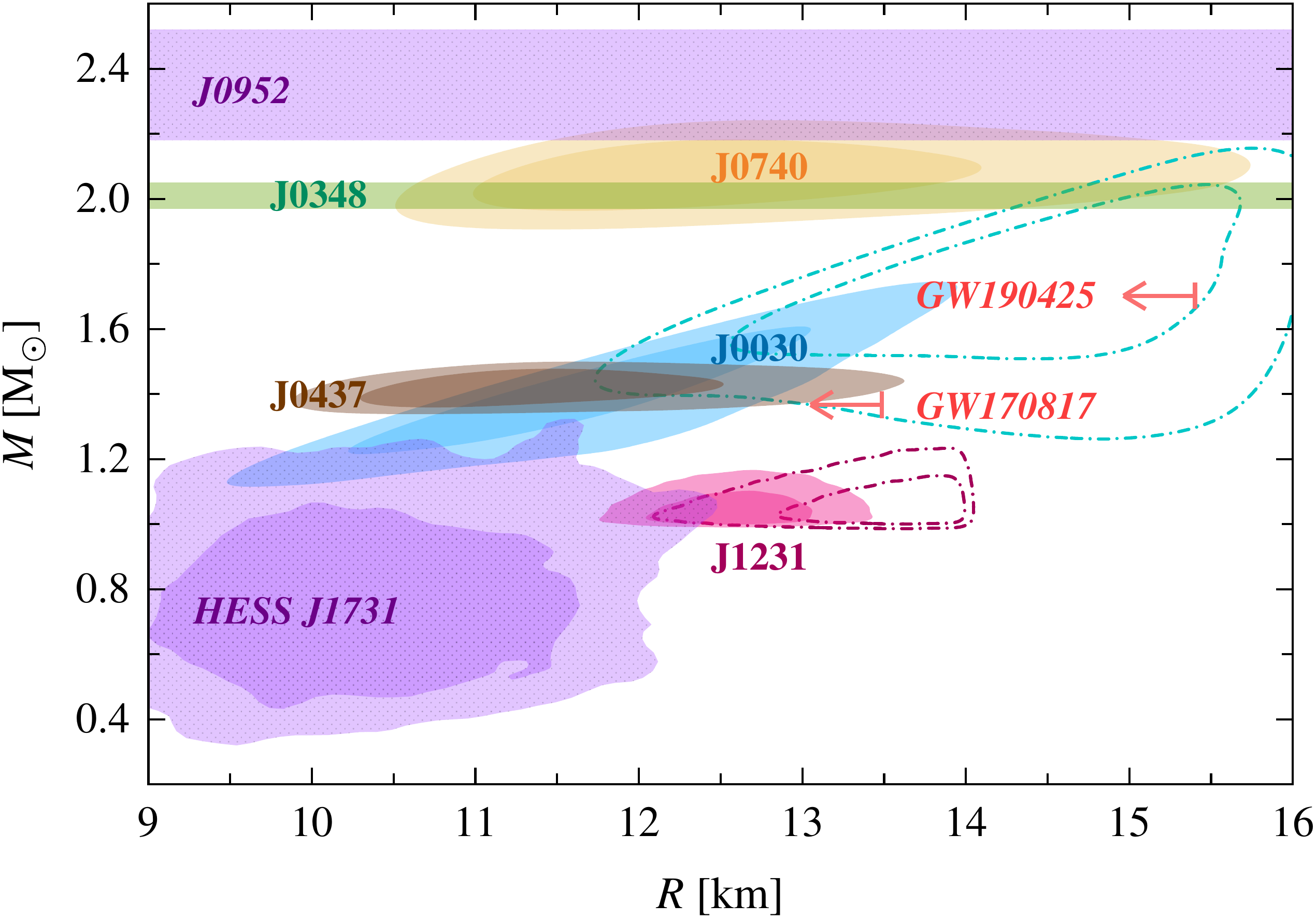}
\caption{ The observational mass and/or radius estimates that 
  represent the multimessenger astrophysical constraints adopted 
  in this study. These ellipses correspond to the 68.3\% and 95.4\%
  CIs for the latest and updated NICER measurements of four pulsars, 
  and the results for HESS J1731, and these horizontal bands correspond 
  to the mass only measurements for two massive pulsars at 68.3\% CI, all 
  labeled in the figure. The NICER mass-radius estimates include the most 
  recent measurement for PSR J0437, two estimates for J1231, the reanalysis 
  of J0740, and two possible solutions for J0030. The radius upper limits 
  obtained from two binary-neutron-star mergers are incorporated into 
  the figure as well. Further details are provided in the text. }
\label{fig:MR_overview}
\end{figure}

The main motivation of this work is to extend our previous study
of covariant density functional (CDF, with linear meson-baryon 
couplings and density-dependent coupling constants) based Bayesian 
inference for compact stars (CSs) with well-established astrophysical 
constraints~\citep{Lijj:2024c} to those that including more extreme 
and less certain constraints. This provides a more complete exploration 
of the parameter space that admits the existence of very massive or 
ultra-compact stars. The astrophysical constraints for the sources 
that we considered in the present analysis are summarized in 
Fig.~\ref{fig:MR_overview}. We aim to assess the compatibility of the 
parameter space over which these models are defined with the recent 
multi-messenger observations of CSs. We stress that previous studies 
of this sort of CDF used simplified functions for density dependence 
of meson-baryon couplings~\citep{Malik:2022a,Malik:2022b,Beznogov:2023,Parmar:2024},
which thus reduced the flexibility of the model.

The remainder of this introduction provides a detailed overview of
current astrophysical constrains in Subsec.~\ref{sec:Astro_review} and
a review of the current state of the art of the dense matter theories
in Subsec.~\ref{sec:CDF_review}. Readers familiar with these topics
can proceed directly to Sec.~\ref{sec:Model} which discusses the
specific CDF model of dense matter used in our analysis. The Bayesian
inference framework adopted in this work and underlying numerical
methods are discussed in Sec.~\ref{sec:Bayesian}. Our analysis is
specific to the cases of the ultra-compact low-mass star HESS J1731-347 and
the high-mass star PSR J0952-0607 in Sec.~\ref{sec:Results} where we
study the impact of these data on the overall picture that emerges
once these are included in the inference analysis.  Our conclusions
and outlook are given in Sec.~\ref{sec:Conclusions}.

\subsection{Summary of astrophysical constraints}
\label{sec:Astro_review} 

Significant advancements in astrophysical observations of CSs over 
the past decade have provided valuable constraints on
dense matter theories. The first observation of gravitational wave
(GW) event GW170817 from the inspiral of a binary neutron stars by
the LIGO-Virgo Collaboration placed constraints on the
dimensionless tidal deformability (TD) of this system,
$\tilde{\Lambda}_{\rm GW170817} \leqslant 720$ at 90\% confidence
interval (hereafter CI), and on the range of the component masses
$1.17$-$1.60\,M_{\odot}$~\citep{LVScientific:2017,LVScientific:2018,LVScientific:2019}.
GW190425 is a highly significant gravitational wave signal, most
likely originating from the merger of two neutron stars, making it the
second such event ever observed. The TD parameter is
constrained to $\tilde{\Lambda}_{\mathrm{GW} 190425} \leqslant 600$,
with component masses being within the range
$1.46$-$1.87\,M_{\odot}$~\citep{LVScientific:2020a}.  Neutron Star
Interior Composition Explorer Mission (NICER) has provided mass–radius
estimates for four pulsars by observing X-ray emission from hot spots
on the neutron stars' surfaces.  The first estimates for one of the
most massive known neutron star PSR
J0740+6620~\citep{Miller:2021,Riley:2021} was updated from a joint
NICER and XMM-Newton analysis of the 2018-2022 dataset, refining the
radius and mass to $12.49^{+1.28}_{-0.88}$~km and
$2.073^{+0.069}_{-0.069}\,M_{\odot}$,
respectively~\citep{Salmi:2024a}. The updated analysis of the
isolated pulsar PSR J0030+0451 revealed that it is a complex source to
model, with solutions varying depending on the assumed model.
Reference~\citep{Vinciguerra:2024} adopted the same models used in the
initial studies~\citep{Miller:2019,Riley:2019}, and found compatible
but slightly more stringent results $R = 13.11^{+1.30}_{-1.30}$~km and
$M = 1.37^{+0.17}_{-0.17}\,M_{\odot}$. When XMM-Newton data are accounted
for, the model that is more consistent with the magnetic field
geometry inferred for the gamma-ray emission for this source inferred radius and mass of $R = 11.71^{+0.88}_{-0.83}$~km and
$M = 1.40^{+0.13}_{-0.12}\,M_{\odot}$ (at 68.3\% CI).  The most complex
model tested in Ref.~\citep{Vinciguerra:2024}, which is highly
preferred by the Bayesian evidence, predicts
$R = 14.438^{+0.883}_{-1.050}$~km and
$M = 1.70^{+0.18}_{-0.19}\,M_{\odot}$. Recently, new NICER
astrophysical constraints have been released for two pulsars --- one
canonical-mass $1.4\,M_{\odot}$ star PSR
J0437-4715~\citep{Choudhury:2024}, and the one-solar-mass star PSR
J1231-1411~\citep{Salmi:2024b}. For the nearest and brightest pulsar
PSR J0437-4715, its first mass and radius are
$M = 1.418^{+0.037}_{-0.037}\,M_{\odot}$ and
$R = 11.36^{+0.95}_{-0.63}$~km, by using the 2017-2021 NICER X-ray
spectral-timing data~\citep{Choudhury:2024}. The inference results for
PSR J1231-1411 exhibit a strong sensitivity to the choice of radius
priors, with stable and well-converged outcomes achieved only under
constrained radius priors. When the radius is restricted to agree with
previous observational constraints and EoS analyses, the inferred
values are $R=12.60_{-0.33}^{+0.31}$~km and
$M=1.04_{-0.03}^{+0.05} M_{\odot}$~\citep{Salmi:2024b}.

The mass range for neutron stars covered by the aforementioned data
spans the range 1-$2\,M_\odot$. Correspondingly, the relevant range
for densities spans $2$-$4\,\rho_{\rm sat}$, with
$\rho_{\rm sat} \simeq 0.16$~fm$^{-3}$ being the nuclear saturation
density. This opens an unprecedented opportunity to explore models of
modern theories of dense matter subject to astrophysical constraints.
For example, the impact of variations in the estimates for PSR
J0030+0451~\citep{Vinciguerra:2024} on dense matter theories and CS
properties have recently been
assessed~\citep{Rutherford:2024,Lijj:2024b,Lijj:2024c}. In addition,
the X-ray observations of the temperature of the neutron star in the
Cassiopeia A supernova remnant and modeling of its thermal evolution
give insights into the possible superfluid properties of the core
matter~\citep{Shternin:2021}.

Determining the maximum mass of neutron stars is crucial for
understanding stellar evolution, supernovae, and binary mergers, as
well as their observational signatures. This value primarily depends
on the EoS at high densities, $\rho > 4 \rho_{\rm{sat}}$. 
Most observed neutron stars rotate significantly more slowly than
their theoretical Keplerian break-up limit (e.g. Ref~\cite{Lijj:2023a} 
and references therein). As a result, mass measurements are commonly 
used to establish lower bounds on the maximum mass of cold, non-rotating 
neutron stars $M_{\rm max}$, even though rapid rotation can support 
up to $\sim 20\%$ additional mass. The highest precisely and reliably 
measured massive mass belongs to pulsar PSR J0348+0432, with a mass 
estimate of $M = 2.01_{-0.04}^{+0.04}\,M_{\odot}$ (at 68.3 \% Cl)~\citep{Antoniadis:2013}.
The heavier pulsar, PSR J0740+6620, has a mass of
$M = 2.08_{-0.07}^{+0.07}\,M_{\odot}$~\citep{NANOGrav:2019,Fonseca:2021} 
and excludes all the EoS models which cannot reach this limit. 
At the same time the analysis of the kilonova AT2017gfo and 
GRB170817A~\citep{Soares-Santos:2017,Villar:2017}, associated with 
GW170817, have helped in designing theoretical models of the post-merger
phase of GW170817 that placed an upper limit on the maximum mass 
$M_{\rm max} \sim 2.3\,M_{\odot}$~\citep{Shibata:2017,Margalit:2017,Ruiz:2018,Rezzolla:2018,Shibata:2019,Khadkikar:2021}.

These mass limits have been further challenged by evidence suggesting
the existence of even heavier compact objects. The ``black widow''
pulsar PSR J0952-0607, the second-fastest known spinning CS
with a rotation frequency of 707 Hz, was detected in the Milky
Way with a mass estimate of $M = 2.35_{-0.17}^{+0.17}\,M_{\odot}$
(at 68.3\% CI)~\citep{Romani:2022}, making it the most massive neutron 
star observed to date.

More recently, pulsar timing observations conducted with the Karoo
Array Telescope (MeerKAT) revealed that the eccentric binary
millisecond pulsar PSR J0514-4002E in the globular cluster NGC 1851
has a companion with a mass in the range of $2.09$-$2.71\,M_{\odot}$ 
(at 95.4\% Cl)~\citep{Barr:2024}. This mass falls within the so-called 
``mass gap'' $(2.5 \lesssim M / M_{\odot} \lesssim 5)$ between known neutron
stars and black holes, suggesting either an exceptionally massive
neutron star or a low-mass black hole.

Additionally, compact objects observed in GW events have been found 
within this mass gap. Examples include the lighter components of black 
hole-black hole or black hole-neutron star merger candidates, such as 
GW190814, where the secondary has a mass of
$M_2 = 2.59_{-0.09}^{+0.08} M_{\odot}$ (at 90\% CI)~\citep{LVScientific:2020b},
and GW200210, with a secondary mass of
$M_2 = 2.83_{-0.42}^{+0.47} M_{\odot}$~\citep{LVKScientific:2023}.
Conversely, GW230529 features a heavier primary component with a mass
of $M_1 = 3.6_{-1.2}^{+0.8} M_{\odot}$ (at 90\% CI)~\citep{LVKScientific:2024},
placing it at the boundary between neutron stars and black holes. The
secondary component of GW190814 has previously been interpreted as the
remnant of an earlier merger that subsequently acquired a more massive
black hole companion through dynamical exchange encounters in a
globular cluster~\citep{LVScientific:2020b}.

In addition to these massive objects, some new results for lower 
mass CSs, which are awaiting confirmation, have been reported.
The X-ray spectrum of the central compact object within the supernova 
remnant HESS J1731-347 was modeled and in combination with the 
distance estimates from Gaia observations yielded a very low mass 
$M=0.77_{-0.17}^{+0.20}\,M_{\odot}$
and a relatively small radius 
$R=10.4_{-0.78}^{+0.86}$~km (at 68.3\% CI)~\citep{Doroshenko:2022}.
According to this estimation, the star can be either the 
lightest neutron star with conventional models of dense matter or a strange
star~\citep{Lijj:2023c,Brodie:2023,Kubis:2023,Sagun:2023,Gao:2024,Lijj:2024a}. 
The analysis of Ref.~\citep{Doroshenko:2022}, however, has been questioned, 
in particular, their use of the object's distance and uniform-temperature 
carbon atmosphere model, see Ref.~\citep{Alford:2023}. It has been also 
argued that such a model does not perform well on longer-exposure 
XMM-Newton data from 2014 for the same source. Instead, a model 
featuring two hot regions emitting blackbody radiation from the surface 
of a star of mass $\sim 1.4\,M_{\odot}$, provides a better description 
of this spectrum~\citep{Alford:2023}. Very recently, the pulse profile 
modeling for the accretion-powered millisecond pulsar XTE J1814-338 
with its thermonuclear burst oscillations, provided puzzling values for 
the mass and radius, $M=1.21_{-0.05}^{+0.05}\,M_{\odot}$ and
$R=7.2_{-0.4}^{+0.3}$~km (at 68.3\% CI)~\citep{Kini:2024}.

\subsection{Summary of nucleonic models of dense matter}
\label{sec:CDF_review}
The EoS for dense matter in CSs
can be constructed using various approaches, including {\it ab initio}
calculations~\citep{Hebeler:2013,Lynn:2016,Drischler:2019,Keller:2023},
density functional
methods~\citep{Oertel:2017,Grasso:2018,Piekarewicz:2020,Sedrakian:2023},
and meta-type models~\citep{Margueron:2018,Zhang:2020,Tsang:2023},
which provide a more schematic representation by parametrizing
nuclear saturation properties. These models can also be linked to
heavy-ion collision data~\citep{Huth:2022,Yao:2024}.

{\it Ab initio} calculations start with the nuclear force constrained by
few-body bound states and scattering phase-shifts, but are
computationally expensive, making it impractical to generate large
ensembles of EoSs covering a broad nuclear parameter space. While
chiral effective field theory ($\chi$EFT) provides a systematic
low-energy expansion, its applicability is currently limited to
densities below $\rho < 2\,\rho_{\rm sat}$. Consequently,
extrapolations to higher densities often rely on physics-agnostic
models~\citep{Hebeler:2013,Annala:2018,Raaijmakers:2019,Rutherford:2024}.

Meta-type models offer computational efficiency but primarily encode
bulk nuclear matter properties, limiting their applicability to finite
nuclei~\citep{Margueron:2018,Zhang:2020,Tsang:2023}. Density functionals, 
on the other hand, provide a balance between computational speed and 
versatility, allowing for EoS calculations while remaining applicable 
to finite nuclear systems~\citep{Oertel:2017,Grasso:2018,Piekarewicz:2020,Sedrakian:2023}.
This makes them particularly well-suited for integrating
nuclear experimental data with astrophysical observations.

The covariant density functional (CDF) models provide a rigorous
framework to address the full range of available data on nuclear
systems, ranging from the atomic chart to the astrophysics of CSs, 
for reviews see Refs.~\citep{Vretenar:2005,Niksic:2011,Oertel:2017,Piekarewicz:2020,Sedrakian:2023}.
These models were instrumental in addressing successfully such
astrophysics problems as the hyperon puzzle in conjunction with two-solar
mass CSs and the TD inference of the GW170817 event. The CDF models are
separated into broad classes which (a) have non-linear meson
contributions to the effective Lagrangian and (b) keep only linear
coupling but impose density-dependence of the coupling, which captures
the medium modifications of the meson-nucleon vertices. The mesons
effectively considered, and their self- or cross-couplings in the
former class or the explicit forms of density-dependences of the
couplings in the latter class, are commonly flexible in order to
address specific issues. Consequently, their predictions, 
are somewhat model-dependent, as different density functionals result
in different dependences on density, see
reviews~\citep{Oertel:2017,Piekarewicz:2020,Sedrakian:2023}.

\section{Density-dependent covariant density functionals for nucleonic matter}
\label{sec:Model}
In this work, we use the standard form of CDF in which Dirac baryons
are coupled to meson fields with density-dependent
couplings~\cite{Typel:1999,Lalazissis:2005,Oertel:2017,Sedrakian:2023,Lijj:2023b}.
The theory is Lorentz invariant and preserves causality when applied
to high-density matter. The baryons interact via exchanges of
$\sigma$, $\omega$, and $\rho$ mesons, which comprise the minimal set
necessary for a quantitative description of nuclear phenomena.

We concentrate on cold, neutrino-free, catalyzed stellar matter.
The Lagrangian with nucleonic degrees of freedom reads
$ \mathscr{L} = \mathscr{L}_N + \mathscr{L}_M + \mathscr{L}_l, $
where the nucleonic Lagrangian is given by
\begin{align}
\label{eq:Lagrangian}
\mathscr{L}_N  = 
\sum_{N = n, p}\bar\psi_N \Big[\gamma^\mu
\big(i \partial_\mu - g_{\omega}\omega_\mu 
    - g_{\rho} \bm{\tau} \cdot \bm{\rho}_\mu \big)
    -\big(m_N - g_{\sigma}\sigma\big) \Big]\psi_N, \nonumber \\
\end{align}
%
where $\psi_N$ are the nucleonic Dirac fields with masses $m_{N}$, 
$\bm{\tau}$ is the vector of isospin Pauli matrices with $\tau_3$ 
being its third component, and $\sigma,\,\omega_\mu$, and 
$\bm{\rho}_\mu$ represent the mesonic fields that mediate the 
interaction among nucleon fields with the corresponding coupling 
constants $g_{\sigma},\,g_{\omega}$ and $g_{\rho}$, respectively. 
The remaining pieces of the Lagrangian correspond to the mesonic, 
and leptonic contributions, 
respectively~\citep{Oertel:2017,Sedrakian:2023}. 

The particle masses adopted in the present analysis are listed in 
Table~\ref{tab:Particle_mass}, according to the widely 
used DD-ME parametrization~\citep{Lalazissis:2005,Lijj:2023b}.
Note that the mass of the $\sigma$ meson $m_\sigma$, which is supposed 
to represent the two-$\pi$-exchange contribution to the nuclear force 
is not known with precision and lies around 500~MeV. The properties
of the infinite nuclear matter described by the present model depend 
only on the ratios of coupling constants and the corresponding meson 
masses~\citep{Walecka:1974,Reinhard:1989}, and our results do not 
change with variation of $\sigma$ meson mass $m_\sigma$, as its 
variations are equivalently encoded into the variations of $g_\sigma$ 
as fixed $m_\sigma$.

\begin{table}[tb]
\centering
\caption{Nucleon and meson masses (in units of MeV) 
adopted in the CDF parametrization.} 
\setlength{\tabcolsep}{13.6pt}
\label{tab:Particle_mass}
\begin{tabular}{cccc}
\hline\hline
$m_N$    & $m_\sigma$ & $m_\omega$ & $m_\rho$ \\
\hline
939.0000 & 550.1238   & 783.0000   & 763.0000 \\
\hline\hline
\end{tabular}
\end{table}

The dependence of meson-nucleon couplings on the total baryon
(vector) density is assumed to be
\begin{align}
g_{m}(\rho)=g_{m}(\rho_{\rm {sat}})f_m(r),
\end{align}
with a constant value at saturation density $g_{m}(\rho_{\rm{sat}})$, 
and a function $f_m(r)$ that depends on the ratio 
$r=\rho/\rho_{\rm{sat}}$. For the isoscalar 
sector, the density dependence is defined as
%
\begin{align}\label{eq:isoscalar_coupling}
f_{m}(r)=a_m\frac{1+b_m(r+d_m)^2}{1+c_m(r+d_m)^2}, \quad m = \sigma,\omega.
\end{align}
%
Typically, parameters in Eq.~\eqref{eq:isoscalar_coupling} are 
not independent because the following five conditions are 
imposed~\citep{Lalazissis:2005,Lijj:2023b}:
%
\begin{align}\label{eq:coupling_constraints}
f_{m}(1)=1, \quad f^{\prime\prime}_{m}(0)=0, \quad
f^{\prime\prime}_{\sigma}(1)=f^{\prime\prime}_{\omega}(1),
\end{align}
%
which reduce the number of free parameters to three, i.e.,
$a_{\sigma}$, $d_{\sigma}$ and $d_{\omega}$. The rational
function~\eqref{eq:isoscalar_coupling} has an advantage in 
contrast to polynomial or exponential forms, because it mimics 
the vertex functionals in Dirac-Brueckner computations that 
were extracted by mapping the self-energies in the local density
approximation~\citep{Typel:1999}. It is applicable over a wide 
density range covered by Dirac-Brueckner theory and converges to 
a constant at high densities.  For the isovector sector, the 
density dependence is taken in an exponential form:
%
\begin{align}\label{eq:isovector_coupling}
f_{\rho}(r) = \text{exp}\,[-a_\rho (r-1)].
\end{align}
%

From Euler–Lagrangian equations, one can deduce the equations of
motion for the constituent fields. In the Hartree mean-field 
approximation, the meson fields are replaced with their 
respective expectation values, $\bar{\sigma} = \langle \sigma \rangle$,
$\bar{\omega} = \langle \omega_0 \rangle$ and 
$\bar{\rho} = \langle \rho_{0,3}\rangle$, which are in fact 
functions of vector or scalar density~\citep{Lijj:2018a,Sedrakian:2023}.
The energy density and pressure of stellar matter can be expressed, 
respectively, as
%
\begin{subequations}\label{eq:energy_pressure}
\begin{align}
\varepsilon = & + \frac{1}{2}m_\sigma^2\bar{\sigma}^2 + 
               \frac{1}{2}m_\omega^2\bar{\omega}^2 + 
               \frac{1}{2}m_\rho^2\bar{\rho}^2 \nonumber \\ 
              & + \frac{1}{2\pi^2}\sum_{N = n,\,p}(2J+1) 
               \int_0^{\infty} dk\,k^2 E^k_{N} f(E^k_{N}) \nonumber \\ 
              & + \frac{1}{\pi^2}\sum_{l = e,\,\mu}
               \int_0^{\infty}dk\,k^2 E^k_l f(E^k_l) \, , \\
P = & - \frac{1}{2}m_\sigma^2\bar{\sigma}^2 + \frac{1}{2}m_\omega^2\bar{\omega}^2  
      + \frac{1}{2}m_\rho^2\bar{\rho}^2 +\rho_N \Sigma_R \nonumber \\
    & + \frac{1}{6\pi^2}\sum_{N = n,\,p}(2J+1)
        \int_0^{\infty}\!\!\! dk \frac{k^4}{E^k_N} f(E^k_N) \nonumber \\
    & + \frac{1}{3\pi^2}\sum_{l = e,\,\mu}
        \int_0^{\infty}\!\!\! dk \frac{k^4}{E^k_l} f(E^k_l)\,, 
\end{align}
\end{subequations}
%
where $2J+1$ is the baryon degeneracy factor, $f$ is the Fermi 
distribution functions, $E^k_N = \sqrt{k^2+ M^{\ast 2}_{\rm D}}$ 
and $E^k_l=\sqrt{k^2+m_l^2}$ are the single-particle energies of 
nucleons and leptons respectively. The lepton mass $m_l$ can be 
taken equal to its free-space value. The effective Dirac mass 
is defined as
\begin{align}
M^\ast_{\rm D} = m_N - g_\sigma\,\bar{\sigma},
\end{align}
which is important for a quantitative description of finite nuclei,
e.g., spin-orbit splitting~\citep{Lijj:2014}. The rearrangement term 
$\Sigma_R$ in Eqs.~\eqref{eq:energy_pressure} arises from the density 
dependence of meson-baryon couplings and is included to ensure 
thermodynamical consistency. It is seen that at zero temperature, the 
energy density and pressure of stellar matter are functions of 
baryonic density only.

It is now clear that if we fix in the Lagrangian~\eqref{eq:Lagrangian}
the nucleon and meson masses to be (or be close to) the ones in the
vacuum then properties of infinite nuclear matter can be computed
uniquely in terms of seven adjustable parameters. These are the three
coupling constants at saturation density
$(g_\sigma,\,g_\omega,\,g_\rho)$, and four parameters
$(a_\sigma,\,d_\sigma,\,d_\omega,\,a_\rho)$ that control their density
dependence.  We further require conditions of weak equilibrium and
charge neutrality that prevail in CSs. Combined with the
field equations for baryons and mesons, they allow one to determine the
equilibrium composition $\rho_N$ and $\rho_l$ at a given baryon number
density and, thus, determine the EoS of stellar matter.

The energy density of isospin asymmetric matter is customarily split 
into an isoscalar term and an isovector term:
%
\begin{align}
\label{eq:isospin_expansion}
\varepsilon \,(\rho, \delta) \simeq 
E_0\,(\rho) + E_{\rm sym}\,(\rho)\,\delta^2 + {\mathcal O}\,(\delta^4)
\end{align}
%
where $\rho= \rho_n + \rho_p$ is the baryonic density, with 
$\rho_{n(p)}$ denoting the neutron (proton) density, 
$\delta = (\rho_{n}-\rho_{p})/\rho$ is the isospin asymmetry, 
and $E_0(\rho)$ and $E_{\rm sym}(\rho)$ are, respectively, 
the energy of symmetric matter and the symmetry energy, 
which at densities close to the saturation $\rho_{\rm sat}$ 
can be further Taylor expanded as 
%
\begin{subequations}\label{eq:Taylor_expansions}
\begin{align}
E_0\,(\rho)
\approx & \,
E_{\rm{sat}} + \frac{1}{2!}K_{\rm{sat}}\,\chi^2
+ \frac{1}{3!}Q_{\rm{sat}}\,\chi^3 
+ \frac{1}{4!}Z_{\rm{sat}}\,\chi^4, \\
E_{\rm sym}\,(\rho)
\approx & \,
J_{\rm{sym}} + L_{\rm{sym}}\,\chi
+ \frac{1}{2!}K_{\rm{sym}}\,\chi^2 
+ \frac{1}{3!}Q_{\rm{sym}}\,\chi^3,
\end{align}
\end{subequations}
%
where $\chi =(\rho-\rho_{\rm{sat}})/3\rho_{\rm{sat}}$.
The coefficients of the expansion are known as
{\it nuclear matter characteristics at saturation density}, 
namely, {\it binding energy} $E_{\rm{sat}}$,
{\it incompressibility} $K_{\rm{sat}}$, the {\it skewness} 
$Q_{\rm{sat}}$ in the isoscalar sector, and the 
{\it symmetry energy} $J_{\rm{sym}}$ and its {\it slope parameter} $L_{\rm{sym}}$ in the isovector 
sector. Given the five macroscopic characteristics together 
with the preassigned values of $\rho_{\rm sat}$ and 
$M^\ast_{\rm D}$, one can determine uniquely the seven 
adjustable parameters of the density functional defined above.

\begin{figure}[tb]
\centering
\includegraphics[width = 0.46\textwidth]{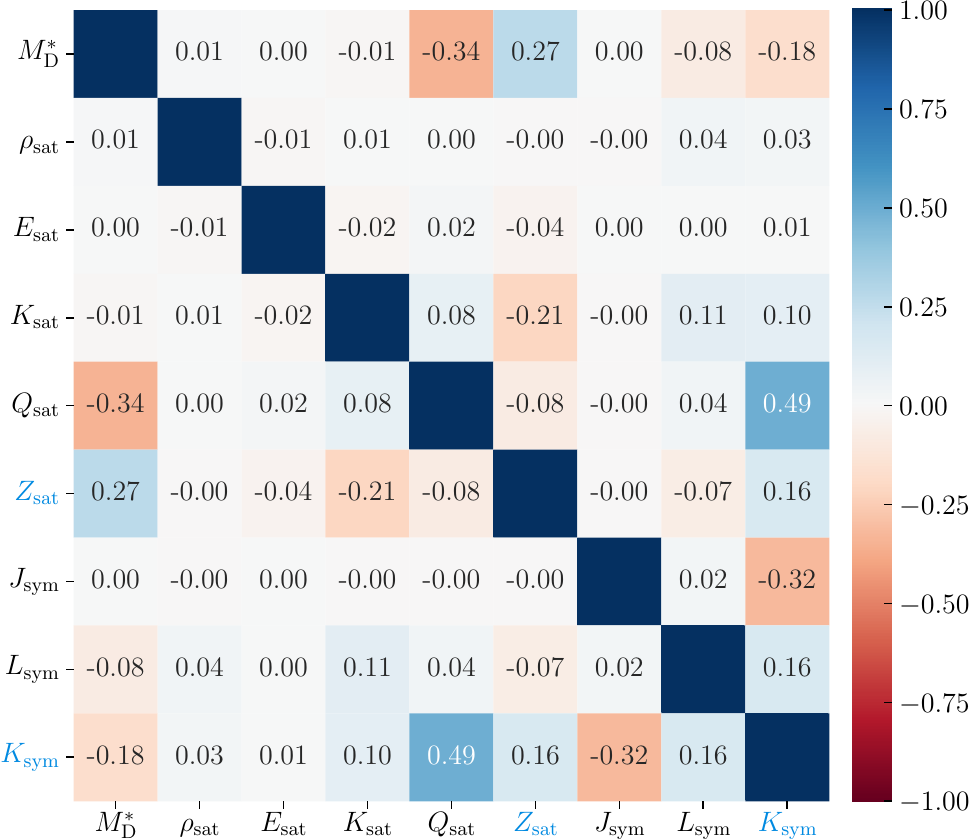}
\caption{ Pearson correlation matrix for variation of nuclear
  characteristic parameters at saturation density, after applying (only) 
  the SNM constraints at saturation density. Note that the higher-order
  parameters $Z_{\rm sat}$ and $K_{\rm sym}$ marked in blue color are
  predictions of the CDF model once the lower-order parameters are
  fixed. }
\label{fig:NM_heatmap}
\end{figure}

\begin{table}[b]
\centering
\caption{ Symmetric nuclear matter (SNM) characteristics at saturation
  density and pure neutron matter (PNM) properties from
  $\chi$EFT. These parameters are used to constrain the CDF
  parameters. The prior distributions assumed for the quantities are
  either a Gaussian distribution (G) or a uniform distribution (U).  }
\setlength{\tabcolsep}{6.2pt}
\label{tab:Nuclear_matter}
\begin{tabular}{cccccc}
\hline\hline
    & No.& Quantity         & Unit         & Interval            & Prior \\
\hline
SNM & 1  & $M_{\rm D}^\ast$ & $m_{\rm N}$  & $  0.60 \pm 0.05$   & G \\
    & 2  & $\rho_{\rm sat}$ & fm$^{-3}$    & $  0.153 \pm 0.005$ & G \\
    & 3  & $E_{\rm sat}$    & MeV          & $-16.1 \pm 0.2$     & G \\
    & 4  & $K_{\rm sat}$    & MeV          & $230 \pm 40$        & G \\
    & 5  & $Q_{\rm sat}$    & MeV          & $[-1000, 1500]$     & U \\
    & 6  & $J_{\rm sym}$    & MeV          & $ 32.5 \pm 2.0$     & G \\
    & 7  & $L_{\rm sym}$    & MeV          & [0, 100]            & U \\
\hline
PNM & 8  & $P\,(\rho)$      &MeV fm$^{-3}$ & N$^3$LO             & G \\
    & 9  &$\epsilon\,(\rho)$&MeV fm$^{-3}$ & N$^3$LO             & G \\
\hline\hline
\end{tabular}
\end{table}
In Eqs.~\eqref{eq:Taylor_expansions}, we truncate the expansion at
different orders in density in the isoscalar and isovector sectors,
while also presenting higher-order coefficients for each
sector—specifically, $Z_{\rm sat}$, $K_{\rm sym}$ and $Q_{\rm sym}$.
These coefficients are, in fact, predictions derived from the
lower-order parameters, because the latter uniquely fix the parameters of
our CDF. The correlation matrix among these coefficients, shown in
Fig.~\ref{fig:NM_heatmap}, is obtained from the posterior distribution
using only the symmetric nuclear matter (SNM) constraints at
saturation density, as listed in the upper panel of
Table~\ref{tab:Nuclear_matter}. The results indicate that most of
these coefficients exhibit little to no correlation, except for the
pair $(M^\ast_{\rm D},\,Q_{\rm sat})$, which still shows a relatively
weak correlation of approximately 0.3 (Pearson coefficient) . Importantly, this framework
allows for the independent variation of specific macroscopic
characteristics within their acceptable ranges while keeping others
fixed. This enables a systematic investigation of how these variations
influence the EoS of dense matter and the properties of 
CSs~\citep{Lijj:2019a,Lijj:2019b}.

\section{Bayesian inference framework}
\label{sec:Bayesian}
Bayesian analysis allows us to infer the probability 
distribution of unknown parameters in a model by 
exploiting the information in observed data. This 
is accomplished via Bayes's theorem,
\begin{align}
P(\bm\theta\vert \bm D) = 
\frac{\mathcal{L}(\bm D\vert\bm\theta)\,P(\bm\theta)}
{\int\mathcal{L}(\bm D\vert\bm\theta)\,P(\bm\theta)\,d\bm\theta},
\end{align}
where available knowledge on model parameters $\bm \theta$ 
is expressed as a prior distribution $P(\bm \theta)$, by 
combining with observational data $\bm D$ in the form of a 
likelihood function $\mathcal{L}(\bm D\vert \bm \theta)$,
the posterior distribution $P(\bm \theta\vert \bm D)$ is 
updated with the information from measured variables. 
The denominator is a normalization factor and acts as 
the evidence of data. 

\begin{table}[b]
\centering
\caption{   The  minimal (``min'') and maximal
(``max") values of the variation range of the CDF parameters at
saturation density over which uniform prior distributions have 
been assigned. 
}
\setlength{\tabcolsep}{20.6pt}
\label{tab:CDF_parameters}
\begin{tabular}{cccc}
\hline\hline
No.& Parameter  & min & max  \\
\hline
1  & $g_\sigma$ & 7.5 & 12.5 \\
2  & $g_\omega$ & 9.5 & 15.5 \\
3  & $g_\rho$   & 2.5 &  5.0 \\
4  & $a_\sigma$ & 1.0 &  2.0 \\
5  & $d_\sigma$ & 0.0 &  1.5 \\
6  & $d_\omega$ & 0.0 &  1.5 \\
7  & $a_\rho$   & 0.0 &  1.5 \\
\hline\hline
\end{tabular}
\end{table}

The present CDF-type EoS is specified by seven adjustable 
parameters in our analysis,
%
\begin{align}
\label{eq:adjustable_parameters_1}
\bm \theta_{\rm EoS} = 
(g_\sigma,\,g_\omega,\,g_\rho,\,a_\sigma,\,d_\sigma,\,d_\omega,\,a_\rho),
\end{align}
%
for which we take uniform prior distributions with the intervals 
listed in Table~\ref{tab:CDF_parameters}. Equivalently, the mapping 
to nuclear matter characteristics
$(M^\ast_{\rm{D}},\,\rho_{\rm{sat}},\,E_{\rm{sat}},\,
K_{\rm{sat}},\,Q_{\rm{sat}},\,J_{\rm{sym}},\,L_{\rm{sym}})$,
allows us to express the gross properties of CSs in terms of 
physically transparent quantities. 

In our process of sampling, for each sample of parameters in
Eq.~\eqref{eq:adjustable_parameters_1} we first calculate the
remaining five non-free parameters
$(b_\sigma,\,c_\sigma,\,a_\omega,\,b_\omega,\,c_\omega)$ in the
isoscalar sector from Eq.~\eqref{eq:coupling_constraints} with a
reference density $\rho_{\rm ref}$. Given these twelve parameters and
$\rho_{\rm ref}$ we can readily compute the seven nuclear matter
characteristics at saturation density listed in
Table~\ref{tab:Nuclear_matter}. The value of $\rho_{\rm ref}$ is
updated each time, simultaneously with the five non-free parameters,
in an iterative process, until the requirement
$\rho_{\rm ref} = \rho_{\rm sat}$ is reached.  At this stage, we have
determined all the relevant parameters for the prior samples. We
then compute the core EoS of stellar matter by incorporating the
additional conditions of weak equilibrium and charge neutrality, 
which govern the composition of CSs. To ensure consistency, we
smoothly match our core EoS to the crust EoS from
Refs.~\citep{Baym:1971a,Baym:1971b} at the crust-core transition
density $0.5\,\rho_{\rm sat}$. Finally, the global properties of
CSs, including mass, radius, and TD, are
obtained by integrating the Tolman-Oppenheimer-Volkoff
equations~\citep{Tolman:1939,Oppenheimer:1939} and the associated
equations for TD~\cite{Flanagan:2008,Hinderer:2008}.

In the present analysis, we employ the Python implementation 
of the Bayesian inference tool 
MULTINEST~\citep{Feroz:2009,Buchner:2014,Feroz:2019}, 
which makes use of the Nested Sampling technique, to generate the 
posterior samples and estimate the marginalized distributions. 
The distributions presented for each run are counted from posterior 
samples of $\sim 3 \times 10^4$ EoSs to ensure the robustness of the 
inferred results. Below, we present a detailed discussion of the 
prior and likelihood we adopted in the current analysis.

\begin{figure}[tb]
\centering
\includegraphics[width = 0.46\textwidth]{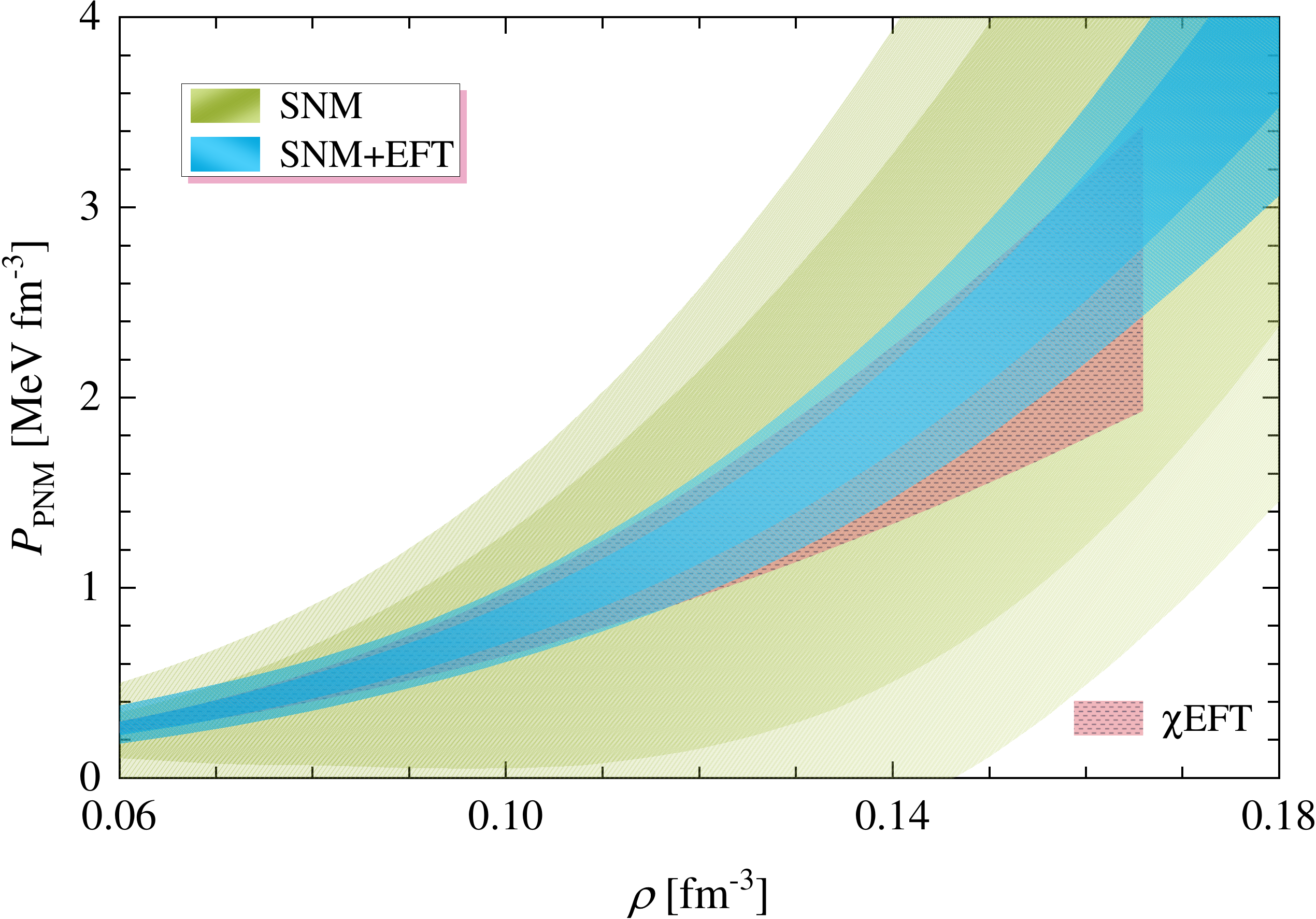}
\caption{ Posterior distributions for the pressure of low-density
  neutron matter at 68.3\% and 95.4\% CIs, after applying either the SNM
  constraints at saturation density only (labeled SNM), or additionally
  the $\chi$EFT constraints (labeled SNM+EFT). }
\label{fig:Pressure_priors}
\end{figure}
%
\begin{figure*}[htb]
\centering
\includegraphics[width = 0.98\textwidth]{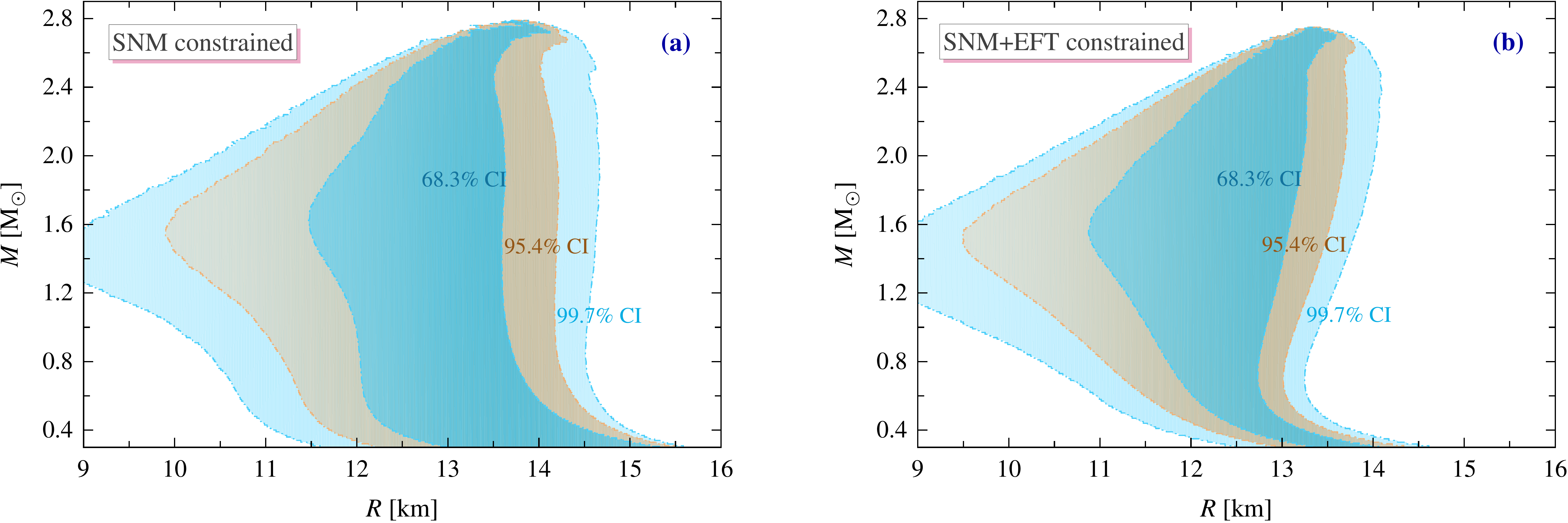}
\caption{ Posteriors for mass-radius distributions at 68.3\%, 95.4\% and 
99.7\% CIs, after applying either the SNM constraints at saturation density 
only (left panel), or additionally the $\chi$EFT constraints (right panel). }
\label{fig:MR_priors}
\end{figure*}
%

\subsection{Nuclear matter properties}
The coefficients of the lower-order terms in the
expansions~\eqref{eq:Taylor_expansions}, specifically $E_{\rm{sat}}$,
$K_{\rm{sat}}$ and $J_{\rm{sym}}$ , are well constrained.
Additionally, the Dirac effective mass $M^\ast_{\rm D}$ at saturation
density --- crucial for an accurate description of finite nuclear
phenomena, including spin-orbit splitting --- is also well constrained.
In contrast, the higher-order coefficients, such as $Q_{\rm{sat}}$ and
$L_{\rm{sym}}$ remain less well determined, though they serve as
asymptotic boundary conditions for the parametrization of CDF
models. For parameters known to within approximately $10\%$, we
assume Gaussian prior distributions, while, for those with greater
uncertainties, we adopt uniform priors. The mean values and standard
deviations, or the respective parameter intervals, are summarized in
Table~\ref{tab:Nuclear_matter}. Notably, we consider broader ranges for 
$Q_{\text {sat}}$ and $L_{\text {sym}}$ to encompass the possible values 
suggested by various works~\citep{PREX:2021,CREX:2022,Reed:2021,Reinhard:2021,Dutra:2014,Margueron:2018,Lijj:2023b}. 

The Gaussian likelihood function for parameter {\it i} is defined as
%
\begin{align}
\mathcal{L}_{ \rm{NM},\,\it{i} }\,(\bm\theta_{\rm{EoS}}) = 
\frac{1}{\sqrt{2\pi}\,\sigma_i}
{\rm exp} \left[-\frac{1}{2}
\left(\frac {d_i-\mu_i(\bm \theta_{\rm{EoS}})}{\sigma_i}\right)^2\right],
\end{align}
%
where $d_i$ and $\sigma_i$ are the datum and its standard 
uncertainty, respectively, and $\mu_i$ the corresponding 
model value. For quantities $Q_{\rm{sat}}$ and $L_{\rm{sym}}$,
the likelihood functions are implemented as Heaviside step
functions of the form
%
\begin{align}
\mathcal{L}_{ \rm{NM},\,\it{i} }\,(\bm\theta_{\rm{EoS}}) = 
H_{\pm}\big[d^{\rm{cut}}_i - \mu_i(\bm \theta_{\rm{EoS}})\big],
\end{align}
%
where the model value $\mu_i$ is constrained from above or
below by the sharp cutoff values $d^{\rm cut}_i$.

In addition to the constraints above, we incorporate results for pure
neutron matter based on chiral effective field theory ($\chi$EFT) 
interactions which constrain the low-density regime of nucleonic EoS. 
In practice, we utilize the N$^3$LO calculations from 
Ref.~\citep{Hebeler:2013} to extract the energy per particle and 
pressure at densities of 0.08, 0.12, and 0.16~fm$^{-3}$, which 
serve as constraints in our analysis. We further assume that the 
associated uncertainties follow a Gaussian distribution at the 
68.3\% CI. This approach allows us to retain models 
that fall outside the strict $\chi$EFT band, rather than discarding 
them outright. It is worth noting that the uncertainty band reported 
in Ref.~\citep{Hebeler:2013} lies within the broader range of results 
obtained from various $\chi$EFT interactions compiled in 
Ref.~\citep{Huth:2022}.

To illustrate the influence of nuclear
matter constraints, Fig.~\ref{fig:Pressure_priors} presents the
pressure distributions for low-density neutron matter at 68.3\% and
95.4\% CIs. These distributions are shown for cases where only the
symmetric nuclear matter (SNM) constraints at saturation density are
applied, as well as when the additional $\chi$EFT constraint is included.
The corresponding mass-radius ($M$-$R$) distributions derived from each
posterior are displayed in Fig.~\ref{fig:MR_priors}. It is evident that
our CDF model can generate a broad range of EoSs, encompassing the
majority of current mass and radius inferences, as summarized in
Fig.~\ref{fig:MR_overview}.

Those CDF parametrizations that are compatible with the $\chi$EFT
are soft at low densities and remain so up to
$2$-$3\,\rho_{\rm sat}$. 
More quantitatively, they have the valus of symmetry energy and its 
slope at nuclear saturation density within the range
$29.5 \leq J_{\rm sym} \leq 33.5$~MeV and $35 \leq L_{\rm sym} \leq 65$~MeV 
(at 95.4\% CI), respectively. 
This leads to small radii for canonical-mass $1.4\,M_{\odot}$ stars, 
$R_{1.4} \lesssim 13.5$~km, as shown in Fig.~\ref{fig:MR_priors}. 
More details on the posteriors of nuclear matter parameters are
discussed in Subsec.~\ref{sec:Results_2}.

\subsection{Astrophysical observations}
We next describe our implementations of the likelihoods for various 
CS measurements, and assume that the datasets of different observed 
stars are independent. 

\subsubsection{NICER data}
To date, the NICER collaborations have delivered the joint 
measurement of mass and radius through pulse profile modeling 
of four millisecond pulsars: a massive $\sim 2.1\,M_{\odot}$ 
star PSR J0740+6620 and two canonical-mass $\sim 1.4\,M_{\odot}$ 
stars PSR J0030+0451 and PSR J0437-4715, and a low-mass 
$\sim 1\,M_{\odot}$ one PSR J1231-1411. 
We construct our likelihood functions for each of the sources 
using the Gaussian kernel density estimation (KDE) 
with the released posterior $(M,\,R)$ samples $\bm S_i$,
\begin{align}
\mathcal{L}_{ \rm{NICER},\,\it{i} }\,(\bm\theta_{\rm{EoS}})= {\rm{KDE}}(M,R|\bm{S}_i),
\end{align}
where the mass $M$ and radius $R$ for the star are functions
of its central pressure and of the sampled EoS parameters. 
For the present analysis, we implement the following measured $(M,\,R)$ 
samples as shown in Fig.~\ref{fig:MR_overview} and listed below. 
The final likelihood is the product of four likelihoods  for each of 
these sources.
\begin{itemize}
\item PSR J0740+6620 (hereafter J0740). 
For this heavy binary pulsar, we incorporate estimates 
from Refs.~\citep{Miller:2019,Riley:2019,Salmi:2024a}. The representative 
estimates used in this study were obtained from a joint NICER and XMM-Newton 
analysis of the 2018-2022 dataset, based on the preferred ST-U model, 
which provides a more comprehensive treatment of the 
background~\citep{Salmi:2024a}.
\item PSR J0437-4715 (J0437). 
We use the first mass and radius estimates for this brightest 
pulsar by using the 2017-2021 NICER X-ray spectral-timing data 
from Ref.~\citep{Choudhury:2024}. The preferred CST+PDT model used 
informative priors on mass, distance, and inclination from PPTA 
radio pulsar timing data, taking into account constraints on the 
non-source background and validating against XMM-Newton 
data~\citep{Choudhury:2024}.
\item PSR J0030+0451 (J0030). 
For this isolated pulsar, we use two alternative mass and radius 
estimates from the reanalysis of 2017-2018 data, as reported in 
Ref.~\citep{Vinciguerra:2024}. The two estimates are based on the 
joint analysis of NICER and XMM-Newton data which are labeled as 
ST+PDT and PDT-U. The ST+PDT results are more consistent with the 
magnetic field geometry inferred for the gamma-ray emission for 
this source~\citep{Kalapotharakos:2021,Vinciguerra:2024}. The PDT-U 
is the most complex model tested in Ref.~\citep{Vinciguerra:2024} 
and is preferred by the Bayesian analysis.
\item PSR J1231-1411 (J1231).
The inference results obtained with the preferred PDT-U model show 
a strong sensitivity to the choice of radius priors, with stable and 
likely converged outcomes achieved only under constrained radius priors.
We consider the mass and radius estimates from the preferred PDT-U 
model with limited radius priors given in Ref.~\citep{Salmi:2024b}. 
One estimate limited the radius to be consistent with the previous 
observational constraints and EoS analyses (model (i)), and the other 
estimate used an uninformative prior for which the radius was limited 
to range 10-14~km (model (ii)).
\end{itemize}
For the sake of completeness, we briefly describe the pulse profile 
modelings mentioned above. In model ST-U each of the two hot spots 
is described by a single spherical cap; in CST+PDT there is a single 
temperature spherical spot with two components, one emitting and one 
masking; in ST+PDT the primary (ST) is described by a single spherical 
cap and the secondary (PDT) by two components, both emitting; in PDT-U 
each of the two hot spots is described by two emitting spherical caps. 
For details see Refs.~\citep{Choudhury:2024,Vinciguerra:2024,Salmi:2024a,Salmi:2024b}.

\begin{table*}[tb]
\centering
\caption{
Astrophysical constraints used for the scenarios in the present work.}
\setlength{\tabcolsep}{9.0pt}
\label{tab:Scenarios}
\centering
\begin{tabular}{ccccccccccccc}
\hline\hline
\multirow{2}*{Scenario}& \multirow{2}*{J0348}& \multirow{2}*{J0952}& GW & J0740 & J0437 & \multicolumn{2}{c}{J0030}& 
\multicolumn{2}{c}{J1231}& HESS \\
\cline{7-8} \cline{9-10}
    &         &         & (2)     & ST-U & CST+PDT & ST+PDT & PDT-U & PDT-U (i) &  PDT-U (ii) & J1731   \\
\hline\hline
 B1 &$\times$ &         & $\times$& $\times$& $\times$& $\times$&         & $\times$&         &         \\
 B2 &$\times$ &         & $\times$& $\times$& $\times$& $\times$&         & $\times$&         & $\times$\\
 B3 &$\times$ & $\times$& $\times$& $\times$& $\times$& $\times$&         & $\times$&         &         \\
 B4 &$\times$ & $\times$& $\times$& $\times$& $\times$& $\times$&         & $\times$&         & $\times$\\
\hline
 F1           &$\times$ &         & $\times$& $\times$& $\times$&         & $\times$&         & $\times$\\
 F2 &$\times$ &         & $\times$& $\times$& $\times$&         & $\times$&         & $\times$& $\times$\\
 F3 &$\times$ & $\times$& $\times$& $\times$& $\times$&         & $\times$&         & $\times$&         \\
 F4 &$\times$ & $\times$& $\times$& $\times$& $\times$&         & $\times$&         & $\times$& $\times$\\
\hline
\hline
\end{tabular}
\end{table*}

\subsubsection{GW data}
The GW170817~\citep{LVScientific:2017}
and GW190425~\citep{LVScientific:2020a} events are the only two confirmed
(and most likely) binary neutron star mergers detected during past
observing runs of the LIGO-Virgo-Kagra collaboration.  When incorporating
measurements from a single GW event, the likelihood function used for
Bayesian inference can be expressed as
\begin{align}
\mathcal{L}(\bm \theta_{\rm{GW}})\varpropto \exp
\Bigg(-2\int\frac{\vert d(f)+h(\bm \theta_{\rm{GW}},f)\vert^2}{S(f)}df\Bigg),
\end{align}
where $S(f)$, $d(f)$, and $h(\bm \theta_{\rm GW},f)$, 
respectively, represent the power spectral density of 
the detector noise, the detected strain signal, and the 
expected strain from a waveform model. The vector 
$\bm \theta_{\rm GW}$ includes parameters that are particularly
useful to infer EoS properties, $\bm \theta^{\rm EoS}_{\rm GW}$, 
and nuisance parameters, $\bm \theta^{\rm nuis.}_{\rm GW}$, that 
are essential when performing analysis on GW-emitting binaries. 
Including a dozen nuisance parameters, however, significantly 
slows down the sampling process and hence we calculate the likelihood 
through the high-precision interpolation in TOAST~\citep{Hernandez:2020}
that marginalizes over these parameters,
\begin{align}
\mathcal{L}_{ \rm{GW},\,\it{i} }\,(\bm\theta_{\rm{EoS}}) = 
F_i\,(\mathcal{M},q,\Lambda_1,\Lambda_2),
\end{align}
where $\mathcal{M} = (M_1 M_2)^{3/5}/(M_1+M_2)^{1/5}$ is the chirp
mass, $q = M_1/M_2$ is the mass ratio of the event $i$, and $\Lambda_1(M_1)$ and
$\Lambda_2(M_2)$ the TDs of the individual stars. 
The TD is related to the component star's mass and radius through the
underlying EoS and its radial profile.

\subsubsection{Massive millisecond pulsars}
To model the mass measurements of massive pulsars (MP), such as PSR
J0348+0432 (hereafter J0948,~\citep{Antoniadis:2013}) and PSR
J0952-0607 (J0952,~\citep{Romani:2022} ), we employ Gaussian distributions. 
The likelihood is then constructed using the cumulative density function of 
the Gaussian distribution
\begin{align}
\mathcal{L}_{\rm{MP},\,\it{i}}\,(\bm\theta_{\rm{EoS}}) = 
\frac{1}{2} \left[1 + {\rm{erf}}
\left(\frac{M_{\rm{max}}(\bm\theta_{\rm{EoS}}) - M_i}
{\sqrt{2}\sigma_i}\right) \right],
\end{align}
where ${\rm erf}\,(x)$ is the error function, $M_i$ and $\sigma_i$ are 
the mean and the standard deviation of the mass measurement for the 
source {\it i}, respectively. We do not take into account the mass 
measurement for PSR J0740~\citep{NANOGrav:2019,Fonseca:2021}, to avoid 
a double counting with its NICER estimates.

\subsubsection{HESS J1731-347}
In the case of unusually light neutron star HESS J1731-347 (hereafter J1731), 
a consensus on its gross parameters has not been reached yet on the basis of 
observational modeling.  Nevertheless, we tentatively incorporate this source 
into our analysis for theoretical exploration. Specifically, we adopt one of 
the two most reliable results from Ref.~\citep{Doroshenko:2022}, which utilized 
the most advanced physical modeling and the most precise distance measurements
available. Similar to the NICER sources, we apply the KDE method to the released 
model samples to generate the posterior distributions, which are then treated as 
the likelihood $\mathcal{L}_{\rm{HESS}}(\bm\theta_{\rm{EoS}})$.

\begin{figure*}[tb]
\centering
\includegraphics[width = 0.98\textwidth]{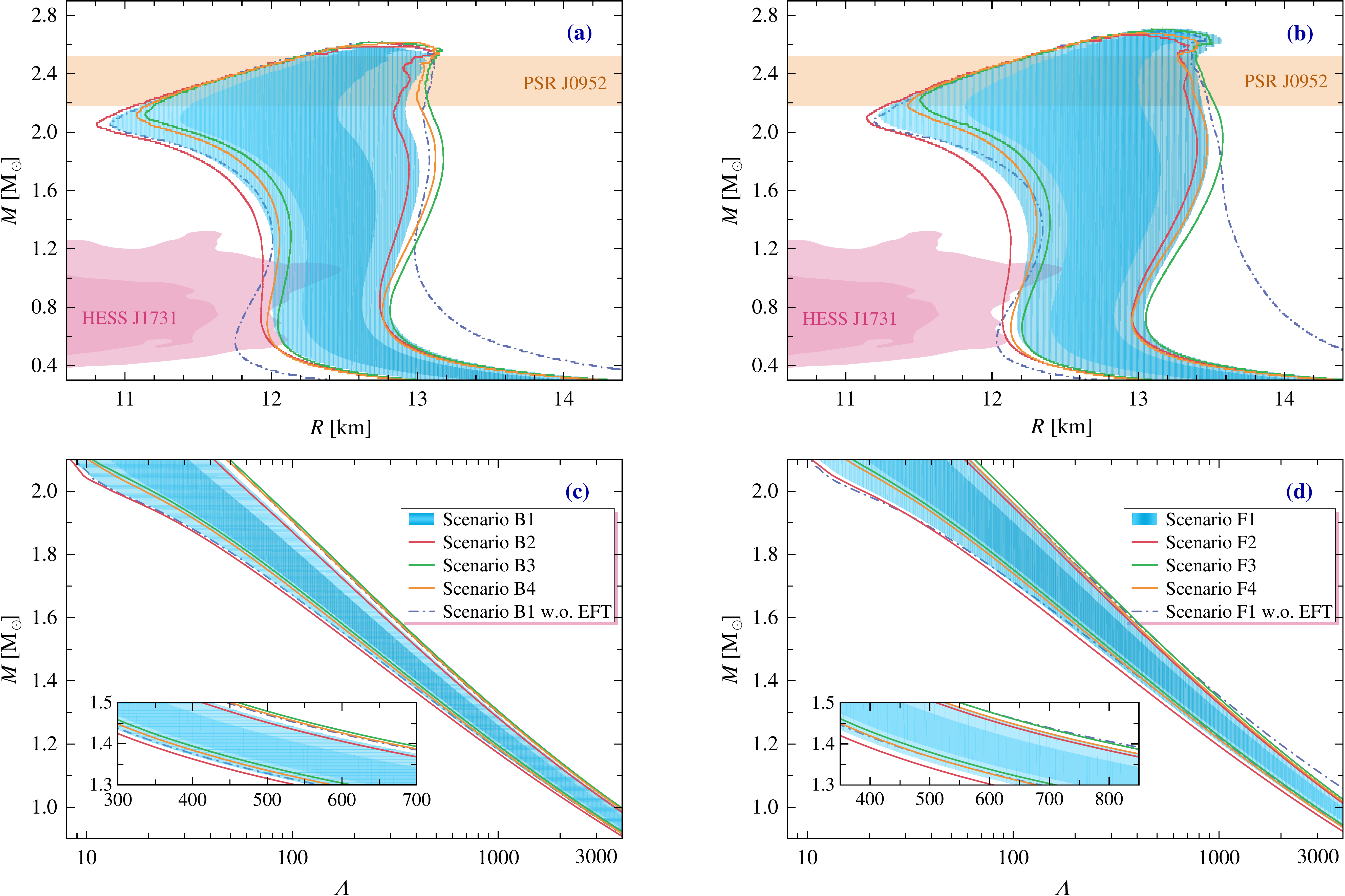}
\caption{ Posteriors for mass-radius (upper panels) and mass-tidal
  deformability (lower panels) distributions under different
  astrophysical scenarios explored in this work. The shaded regions
  represent the distributions at 68.3\% and 95.4\% CIs for scenarios B
  (left panels) and F (right panels). The various color solid lines show 
  the results at 95.4\% CI for scenarios which take additionally into 
  account the mass-radius estimates of HESS J1731 (B2 and F2), the mass
  measurement of PSR J0952 (B3 and F3), or both of these masurements
  (B4 and F4). Finally, the dash-dotted lines correspond to the
  scenarios which do not include the $\chi$EFT constraints.  In the
  lower panels, the insets magnify the TDs for typical masses of CSs
  involved in GW events. In the upper panels, the elliptical contours
  of the input mass-radius estimates of HESS J1731 and the band for the
  mass of PSR J0952 are shown as well for reference. }
\label{fig:MR_scenarios}
\end{figure*}

\begin{figure*}[htb]
\centering
\includegraphics[width = 0.98\textwidth]{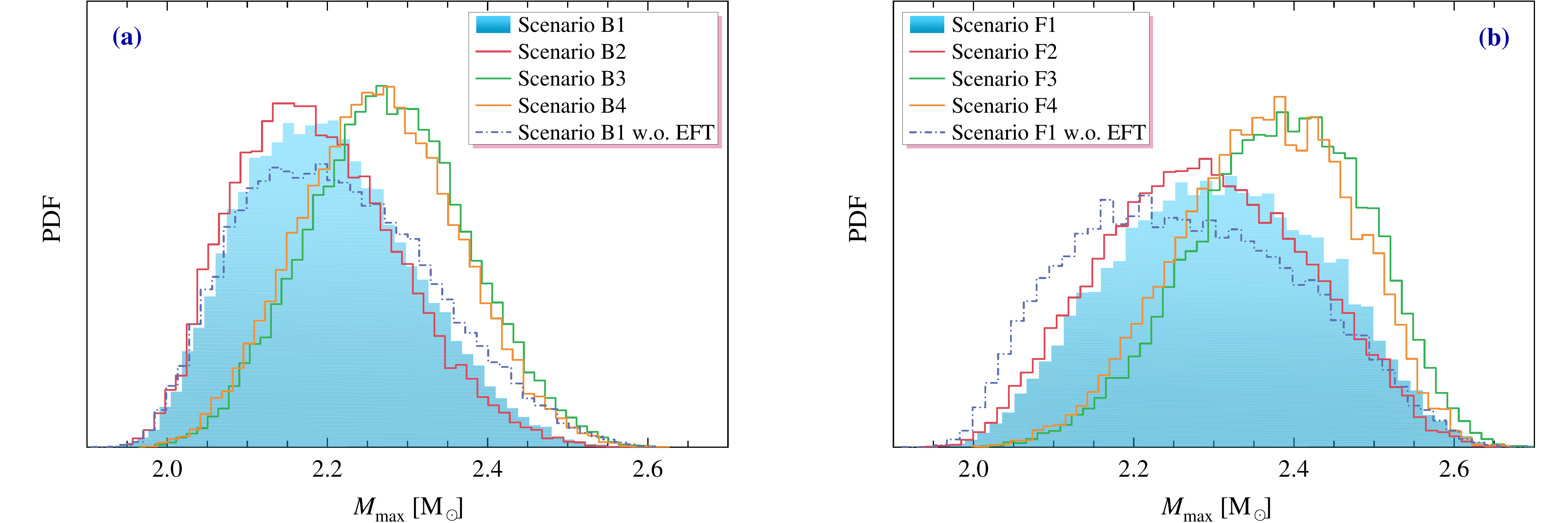}
\caption{Probability distribution functions (PDFs) for maximum mass
  $M_{\rm max}$ under different astrophysical scenarios. The shaded
  regions represent the distributions for scenarios B1 (left panel)
  and F1 (right panel), while the  various color solid lines correspond to the scenarios which take additionally into account the mass-radius estimates of HESS J1731 (B2 and F2), mass measurement of PSR J0952 (B3
  and F3), or both of these two measurements (B4 and F4). The
  dash-dotted lines correspond to neglecting
  the $\chi$EFT constraints in scenarios B1 and F1.}
\label{fig:Mmax_scenarios}
\end{figure*}

\begin{figure*}[htb]
\centering
\includegraphics[width = 0.98\textwidth]{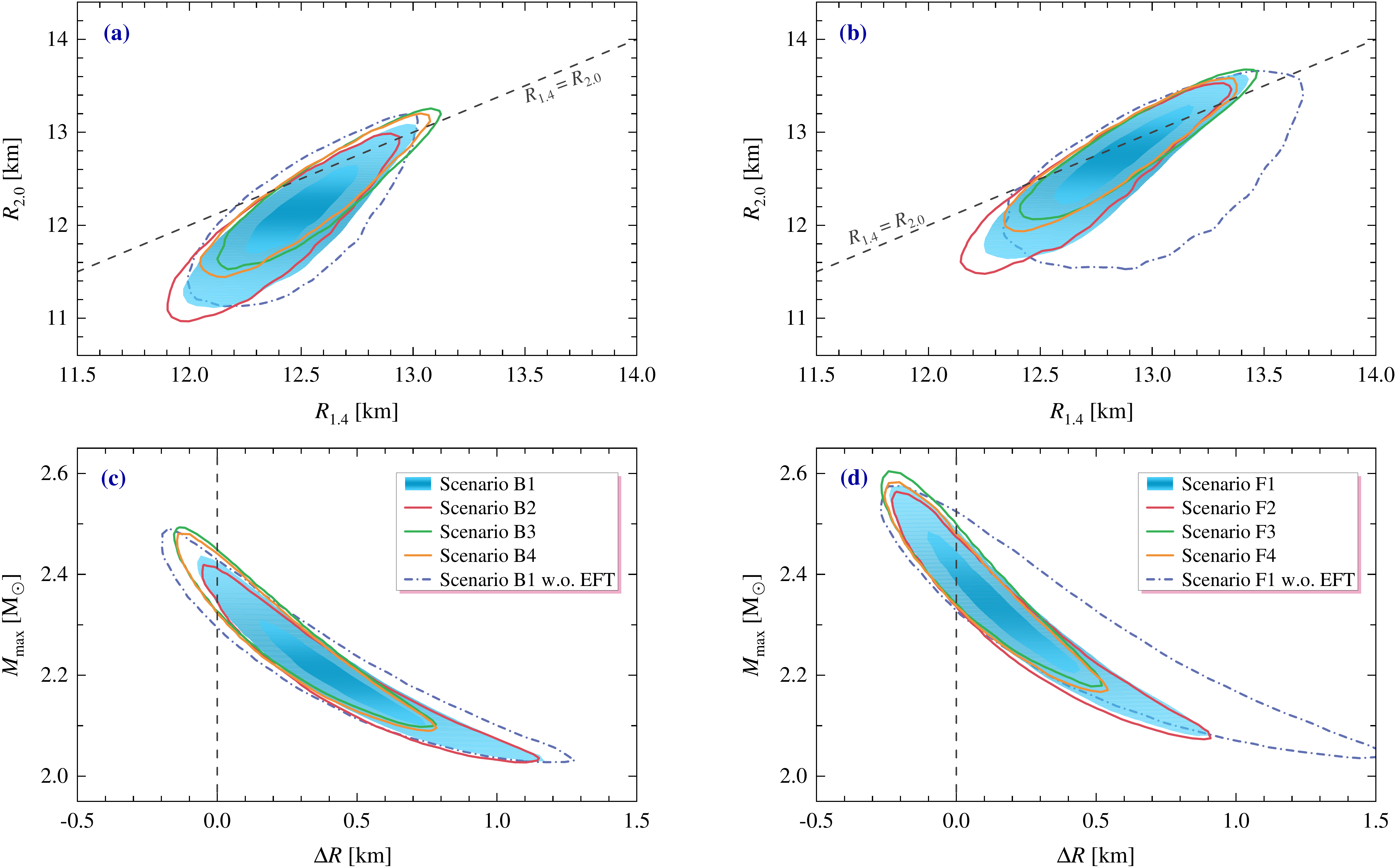}
\caption{
  Correlation between the radii $R_{2.0}$ and $R_{1.4}$ (upper panels), 
  and the maximum mass $M_{\rm max}$ and the 
  radius-difference $\Delta R = R_{1.4}-R_{2.0}$ (lower panels) 
  under different astrophysical scenarios. The shaded regions represent 
  the distributions at 68.3\% and 95.4\% CIs for scenarios B1 (left panels)
  and F1 (right panels), while the various color solid lines correspond to 
  95.4\% CI for the scenarios which take additionally into account the mass-radius estimates of HESS J1731 (B2 and F2), mass measurement of PSR J0952 (B3 and F3), or both of these two measurements (B4 and F4). 
  The dash-dotted line for scenarios (B1 and F1) correspond to neglecting
  the $\chi$EFT constraints.
}
\label{fig:MR_correlations}
\end{figure*}

\section{Multimessenger inferences on compact stars and dense matter}
\label{sec:Results}
We apply the theoretical and astrophysical constraints both
selectively and in concert to elucidate the importance of the
individual impacts of the ultra-compact low-mass star (HESS J1731) and
the very high mass star (PSR J0952) as well as the overall picture
that emerges from a combination of such constraints.

Before embarking onto detailed discussion we present the scenarios
that we consider in Table~\ref{tab:Scenarios}.  Specifically, we will
examine two baseline scenarios, labeled B1 and F1, which share the
same data for PSR J0437 and PSR J0740 but incorporate two different
analyses for PSR J0030 using different models of the surface
temperature patterns, and two different analyses for PSR J1231 using
different radius priors. For each baseline scenario, we implement
further the data for the ultra-compact low-mass star HESS J1731
(denoted as scenarios B2 and F2), and for the high-mass star PSR J0952
(denoted as scenarios B3 and F3). Finally we take into account both
constraints (denoted as scenarios B4 and F4) in our analysis.
 
\subsection{Implications for properties of compact stars}
\label{sec:Results_1}
The new NICER estimates for PSR J0437 and J1231, combined with the
reanalysis of PSR J0030, have enabled the construction of various
scenarios incorporating different models of surface temperature
distributions for pulsars. The scenarios B1 and F1 in
Table~\ref{tab:Scenarios} represent two distinct combinations of the
current NICER estimates for pulsars, corresponding to the softest and
stiffest models, respectively; see also Ref.~\citep{Lijj:2024c}.  In
scenario B1, the more compact estimates for PSR J1231 (PDT-U (i)
model) and J0030 (ST+PDT model), favoring softer EoS at densities
below 2 and $3\,\rho_{\rm sat}$, respectively, are used and give rise
to the tightest credible regions. In contrast, the widest credible
regions are predicted in scenario F1, which incorporates less compact
estimates for PSR J1231 (PDT-U (ii) model) and J0030 (PDT-U model).

Figure~\ref{fig:MR_scenarios} shows the posteriors for $M$-$R$
and $M$-$\Lambda$ distributions within the scenarios B1 and F1, and 
Fig.~\ref{fig:Mmax_scenarios} displays the corresponding probability 
distributions for the values of maximum mass $M_{\rm max}$. 
For reference, we list here the selected gross properties of CSs
predicted from the baseline scenarios B1 and F1. The radius for 
a canonical-mass $1.4\,M_{\odot}$ star is $R_{1.4} = 12.47_{-0.50}^{+0.48}$ 
(at 95.4\% CI) for scenario B1 and $12.79_{-0.56}^{+0.55}$ for 
scenario F1. The corresponding TD is $\Lambda_{1.4} = 472_{-121}^{+163}$ 
for scenario B1 and $571_{-162}^{+207}$ (at 95.4\% CI) for scenario F1. 
The maximum mass is $M_{\rm max} = 2.20_{-0.17}^{+0.23}\,M_{\odot}$ 
and $2.32_{-0.24}^{+0.24}\,M_{\odot}$ respectively for scenarios B1 
and F1. The latter scenario allows for a static CS interpretation for 
the secondary component of the GW190814 event~\citep{LVScientific:2020a}, 
with a mass $2.50$-$2.67\,M_{\odot}$ at 90\% CI, and for the primary 
component of the GW230529 event~\cite{LVKScientific:2024}, with a 
mass $3.6^{+0.8}_{-1.2}\,M_{\odot}$.

With additional constraints incorporated, the $M$-$R$ distribution
retains a similar overall structure. Notably, the 95.4\% CI
regions reveal remarkably narrow radius intervals for sub-canonical
mass CSs ($M < 1.4\,M_{\odot}$), typically around 1~km. In
contrast, the widest radius variations are observed around
$M \thickapprox 2.0\,M_{\odot}$.

\begin{figure*}[htb]
\centering
\includegraphics[width = 0.98\textwidth]{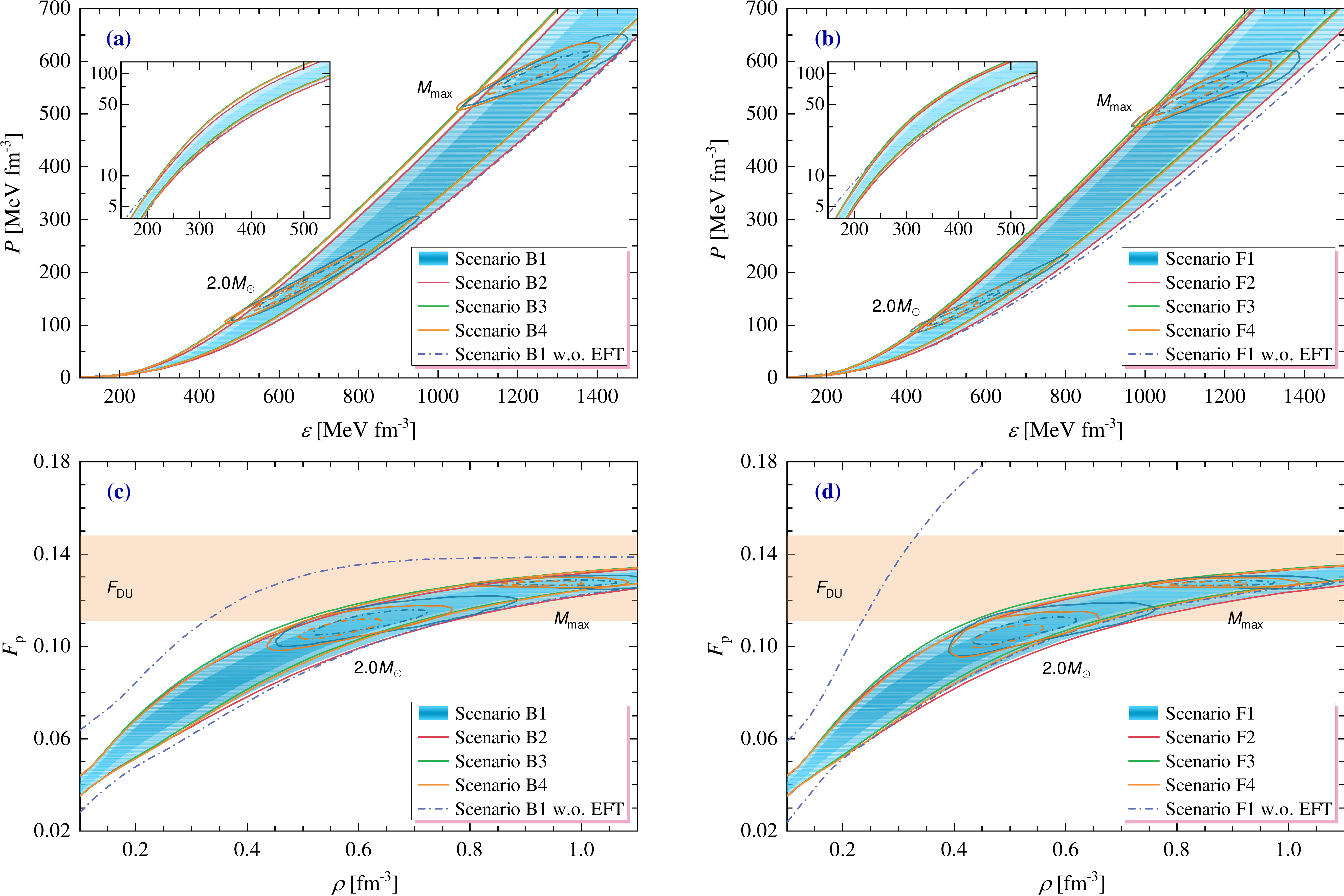}
\caption{ Posteriors for the nucleonic compact-star EoS (upper panels)
  distribution and the proton fraction (lower panels) under different
  scenarios. The shaded regions represent the distributions at 68.3\% 
  and 95.4\% CIs for scenarios B1 (left panels) and F1 (right panels), 
  while the various color solid lines correspond to 95.4\% CI for the 
  scenarios which take additionally into account 
  the mass-radius estimates of HESS J1731
  (B2 and F2), the mass measurement 
  of PSR J0952 (B3 and F3), and both of these two measurements (B4 and F4). The 
  dash-dotted line for scenarios (B1 and F1) corresponds to neglecting 
  the $\chi$EFT constraint. In each panel, the contours show the 
  corresponding distributions of the respective $2.0\,M_{\odot}$ and 
  the maximum-mass configurations for scenarios B1 (B4) and F1 (F4). 
  In lower panels the orange bands labeled $F_{\rm DU}$ show the 
  admissible threshold values for the onset of direct Urca (DU) cooling 
  process.}
\label{fig:EoS_scenarios}
\end{figure*}

\begin{figure*}[htb]
\centering
\includegraphics[width = 0.98\textwidth]{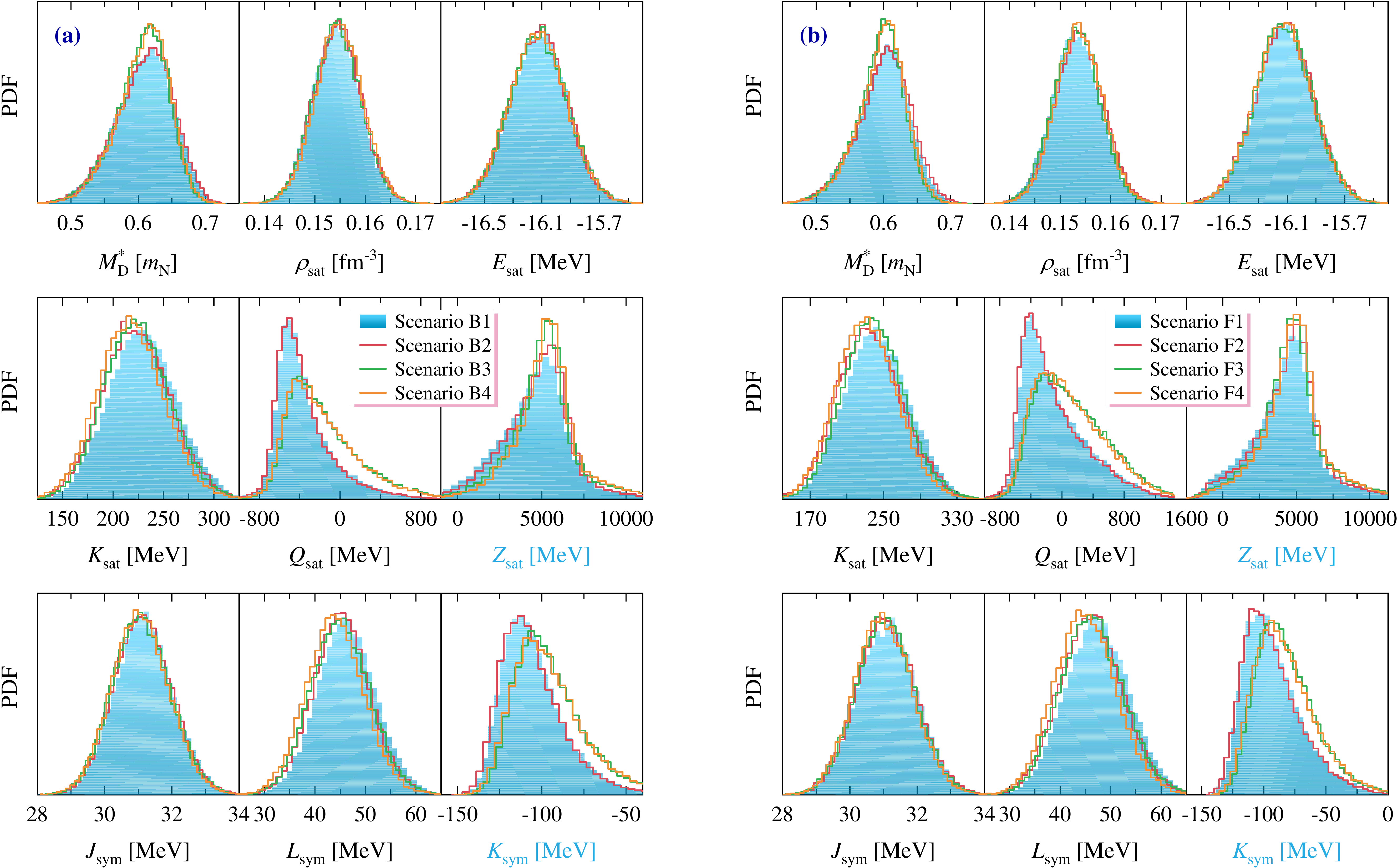}
\caption{ Posterior distributions of nuclear matter saturation
  properties under different astrophysical scenarios. The shaded
  regions indicate the results for scenarios B1 
  (left panels) and F1 (right panels). The various color lines 
  represent the results for scenarios that additionally incorporate
  the mass-radius estimates of HESS J1731 (B2 and F2), the mass
  measurement of PSR J0952 (B3 and F3), or both measurements combined
  (B4 and F4). Note that the higher-order characteristic parameters
  $Z_{\rm sat}$ and $K_{\rm sym}$ marked in blue color are predictions
  of those lower-order parameters in the present CDF. }
\label{fig:NM_characteristics}
\end{figure*}

\begin{figure*}[htb]
\centering
\includegraphics[width = 0.98\textwidth]{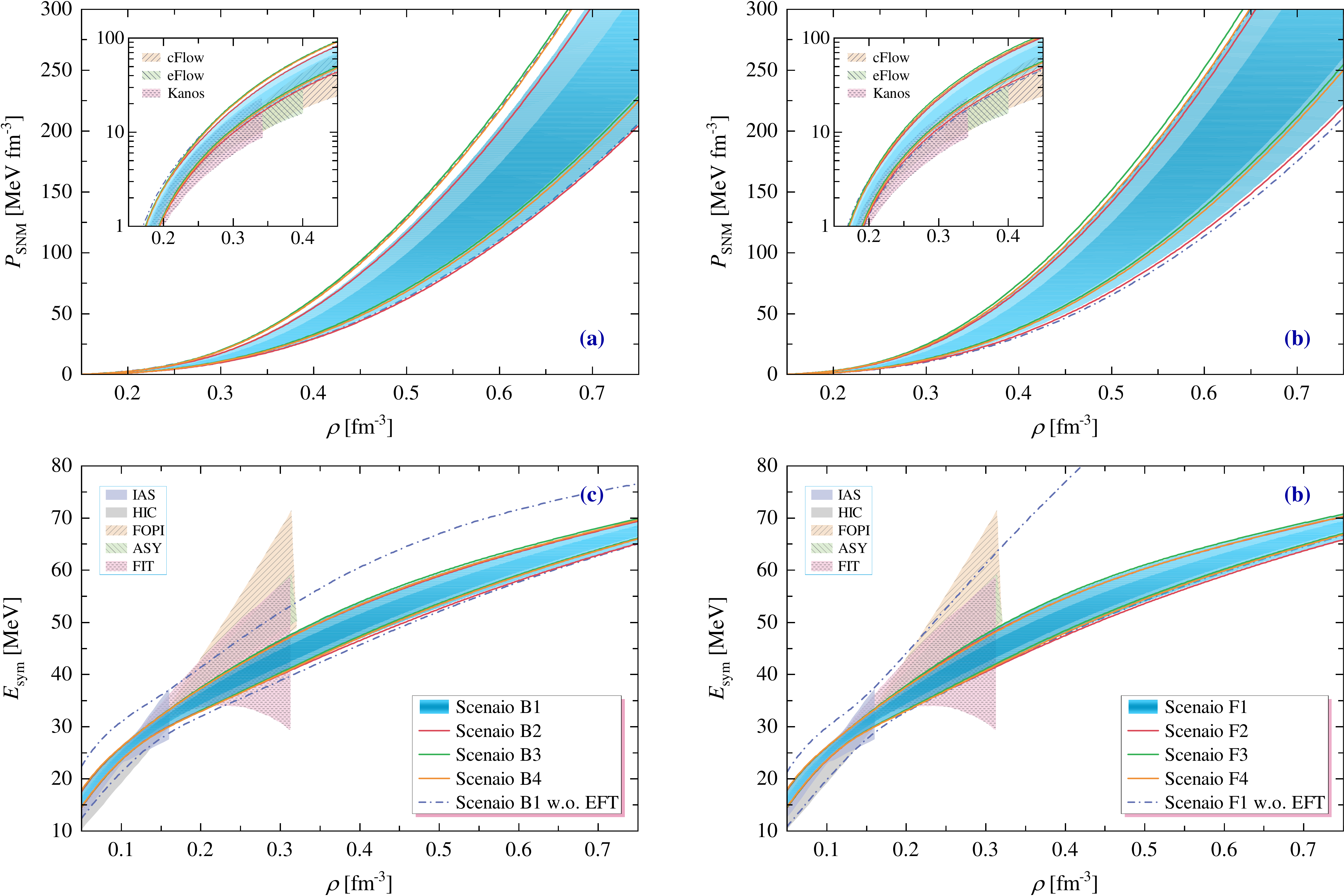}
\caption{ Posterior distributions of pressure (upper panels) and
  symmetry energy (lower panels) as functions of baryonic density for
  symmetric nuclear matter under different astrophysical scenarios. 
  The shaded regions indicate the 68.3\% and 95.4\% CIs for scenarios 
  B1 (left panels) and F1 (right panels). Solid lines represent results 
  at 95.4\% CI incorporating additional constraints from the mass-radius 
  estimates of HESS J1731 (B2 and F2), the mass measurement of PSR J0952 
  (B3 and F3), or both (B4 and F4). Dash-dotted lines correspond to 
  scenarios where the $\chi$EFT constraint is not applied. In the upper 
  panels, the insets magnify the low-density regions. The constraints 
  for pressure and symmetry energy obtained from hadronic transport 
  simulations and/or nuclear structure studies are shown for comparison, 
  see text for details. }
\label{fig:EoS_nuclear}
\end{figure*}

Incorporating the mass and radius estimates for HESS J1731 results in only 
a slight shift—approximately 0.1 km—to the lower-radius side in the
posterior region for scenario B2 (F2) compared to B1 (F1); see
Fig.~\ref{fig:MR_scenarios}. As expected, the posterior distributions
for tidal deformability ($\Lambda$) exhibit similar trends to the
$M$-$R$ posteriors due to the strong positive correlation between
$\Lambda$ and $R$ for a given mass (e.g., Ref~\cite{Lijj:2023a} 
and references therein). Consequently, the posterior region overlaps 
with the mass and radius estimates for HESS J1731 only at the 95.4\%
CI. Meanwhile, the impact on the maximum mass 
($M_{\rm max}$) distribution remains negligible in scenarios B1 and B2
(F1 and F2); see Fig.~\ref{fig:Mmax_scenarios}. This suggests that the
mass and radius measurements of four pulsars from NICER, combined with 
the TDs inferred from GW170817, GW190425 and the constraints from
$\chi$EFT calculations, primarily define the posterior region of the
$M$-$R$ distribution. The effect of the $\chi$EFT constraint on the
$M$-$R$ posterior is particularly evident in scenarios B1 and F1
(Fig.~\ref{fig:MR_scenarios}), where it imposes a lower bound on 
the radii of sub-canonical-mass stars. Additionally, the mass and radius 
inferences derived for HESS J1731 shows only weak overlap with the 
$M$-$R$ ellipse for PSR J1231 (Fig.~\ref{fig:MR_overview}), which 
limits the influence of HESS J1731 data in distinguishing viable EoSs.
 
Including the mass measurement for PSR J0952 instead, namely for
scenario B3 (F3), the posterior region is shifted to a larger radius
region by 0.1-0.4~km compared to scenario B1 (F1). A more significant
difference appears for stars with mass $M > 1.4\,M_{\odot}$. This is
because the radii for sub-canonical-mass stars are well-constrained,
while the NICER inferences for PSR J0030 and J0740, allow for radii
larger than $14$~km, which then allow stiffer high-density EoSs. 
As anticipated, the maximum mass $M_{\rm max}$ is enlarged by
$0.1\,M_{\odot}$, with, in particular, a higher probability density
for scenario F3; see Fig.~\ref{fig:Mmax_scenarios}. The maximum mass
now is $M_{\rm max} = 2.28_{-0.19}^{+0.20}\,M_{\odot}$ and
$2.39_{-0.23}^{+0.19}\,M_{\odot}$ respectively for scenarios B3 and
F3. Meanwhile, the distributions of maximum mass $M_{\rm max}$ is
somewhat narrowed compared to scenarios B1 and F1, respectively.

Finally, incorporating both HESS J1731 and PSR J0952 data, namely 
for scenarios B4 and F4, the boundaries for the posterior regions 
are located well within the ranges set by the two cases discussed
above. The distributions of maximum mass $M_{\rm max}$ are rather
close to those of scenarios B3 and F3, respectively. To conclude, 
the $M$-$R$ estimates for HESS J1731 and mass measurement for PSR J0952
affect only the finer details of the posterior compared to the case
where these constraints are not included. 

Reference~\cite{Drischler:2021} pointed out that the sign of the 
difference of the radii of CSs of masses 1.4 and $2.0\,M_{\odot}$, 
$\Delta R = R_{1.4}-R_{2.0}$ is an indicator that the EoS softens 
(if positive) or stiffens (if negative) at high densities. 
Figure~\ref{fig:MR_correlations} illustrates the posterior distributions 
for the correlation between $R_{1.4}$ and $R_{2.0}$, and the correlation 
between the maximum mass $M_{\rm max}$ and the radius difference 
$\Delta R$. It is seen that the correlations between these parameters 
are similar across all scenarios. For scenarios B (B refers to B1--B4), 
the preference is for $\Delta R$ being positive, i.e., $R_{1.4} > R_{2.0}$, but for scenarios F, negative $\Delta R$ is allowed even at 68.3\% CI.
More details of correlations among properties of CSs will be discussed 
below in Subsec.~\ref{sec:Results_3}. The predicted key quantities of 
CSs under different astrophysical scenarios are summarized in the 
Appendix.

It is worth noting that across all the scenarios discussed, our CDF-based
EoS not only maintains reasonable nuclear saturation properties (to be
elaborated in the next subsection) but also supports maximum masses
exceeding $\approx 2.5\,M_{\odot}$. 
This contrasts with earlier studies of this type, which employed 
simplified forms of the density-dependent meson-baryon coupling 
functions~\eqref{eq:isoscalar_coupling}
~\cite{Malik:2022a,Beznogov:2023,Parmar:2024}, offering less flexibility 
than the approach adopted in the present work. In fact, calculations based 
on the same CDF framework used here~\citep{Providencia:2023,Char:2025}, 
or on versions of CDF with more tunable parameters (beyond simple exponentials)~\citep{Providencia:2023,Char:2023,Scurto:2024,Char:2025}, 
have been able to construct models with $M \geqslant 2.5\,M_{\odot}$, 
even when fewer observational constraints were applied.

\subsection{Implications for properties of nucleonic EoS}
\label{sec:Results_2}

\begin{figure*}[tb]
\centering
\includegraphics[width = 0.98\textwidth]{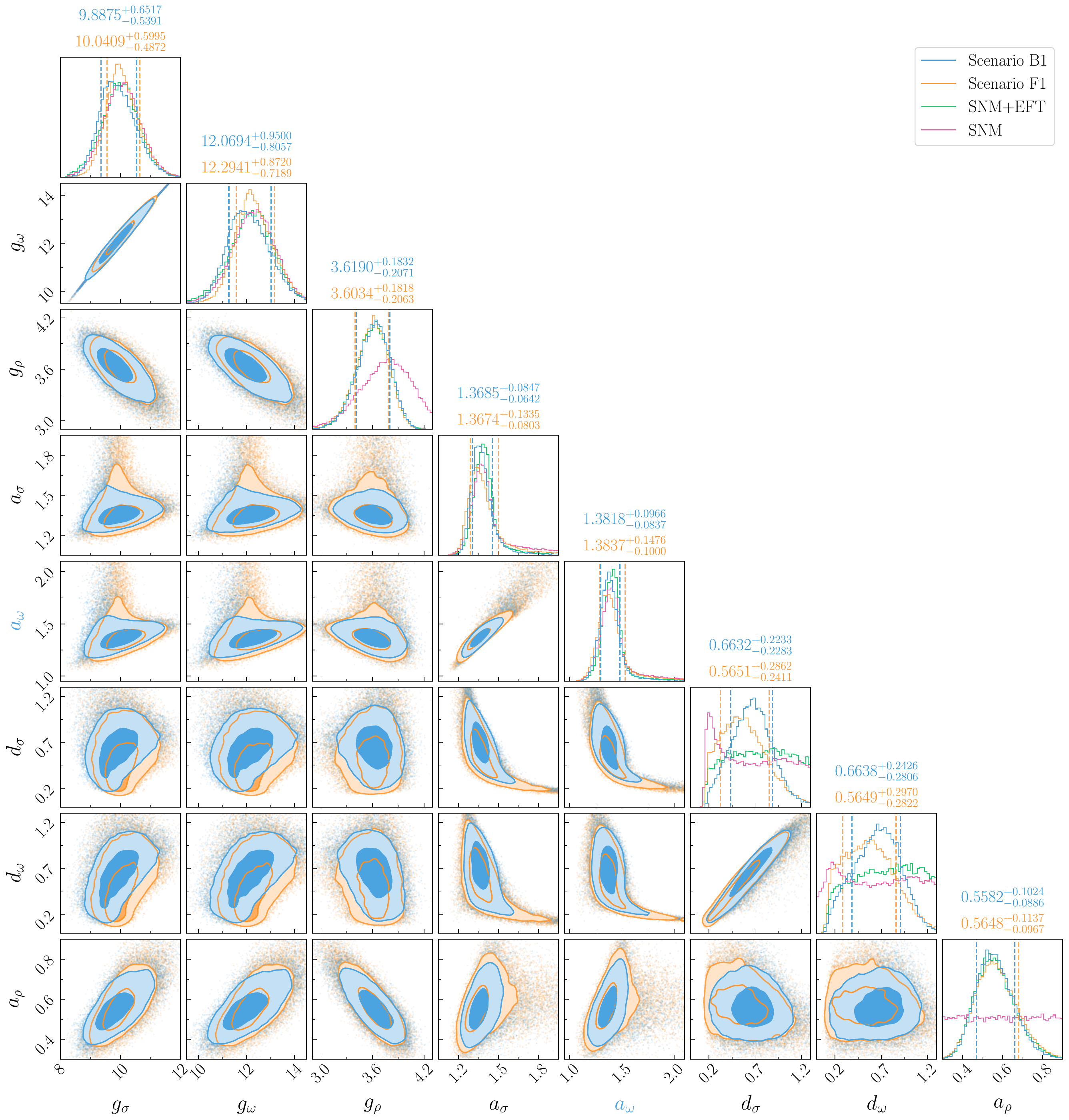}
\caption{ Posterior distributions of eight CDF parameters under
  astrophysical scenarios B1 and F1. The light- and dark-shaded
  regions represent the 68.3\% and 95.4\% CIs for the two-dimensional 
  distributions, while the one-dimensional posteriors for each parameter 
  are shown along the edges, with vertical lines indicating the 68.3\% 
  CI. For comparison, results considering only the constraints 
  from symmetric nuclear matter characteristic parameters (labeled SNM) and 
  those including additional EFT computations (labeled SNM+EFT) are provided in the one-dimensional plots. 
  The parameter $a_{\omega}$ (highlighted in blue) is not a free parameter 
  in the present CDF framework.}
\textsc{\label{fig:CDF_parameters}}
\end{figure*}

\begin{figure*}[tb]
\centering
\includegraphics[width = 0.98\textwidth]{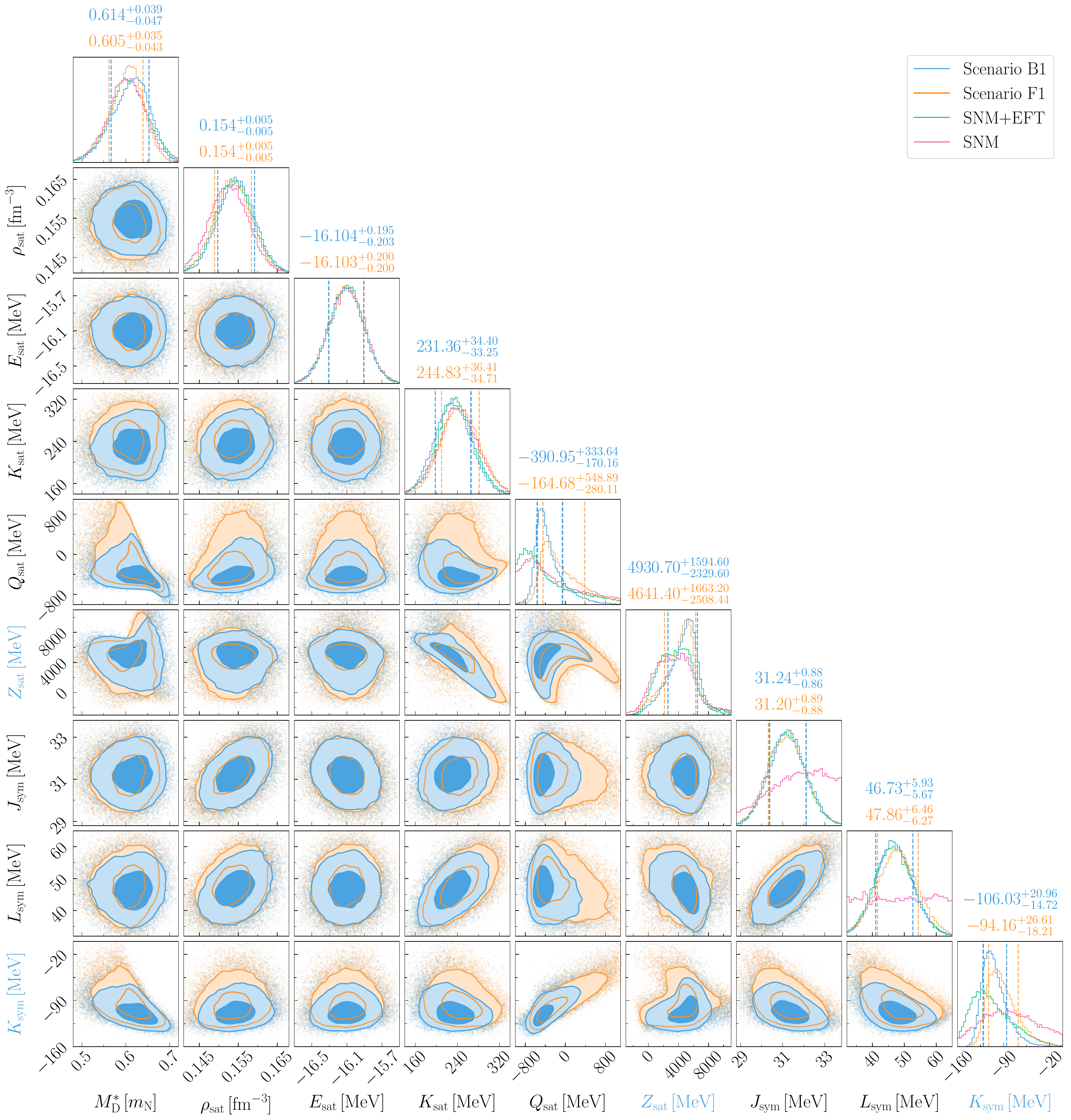}
\caption{
Same as Fig.~\ref{fig:CDF_parameters}, but for nine nuclear 
characteristic parameters at saturation density. Note that the 
higher-order characteristic parameters $Z_{\rm sat}$ and 
$K_{\rm sym}$ marked in blue color are predictions of the CDF 
model, once the lower-order parameters are fixed.
}
\label{fig:NM_properties}
\end{figure*}

\begin{figure*}[tb]
\centering
\includegraphics[width = 0.98\textwidth]{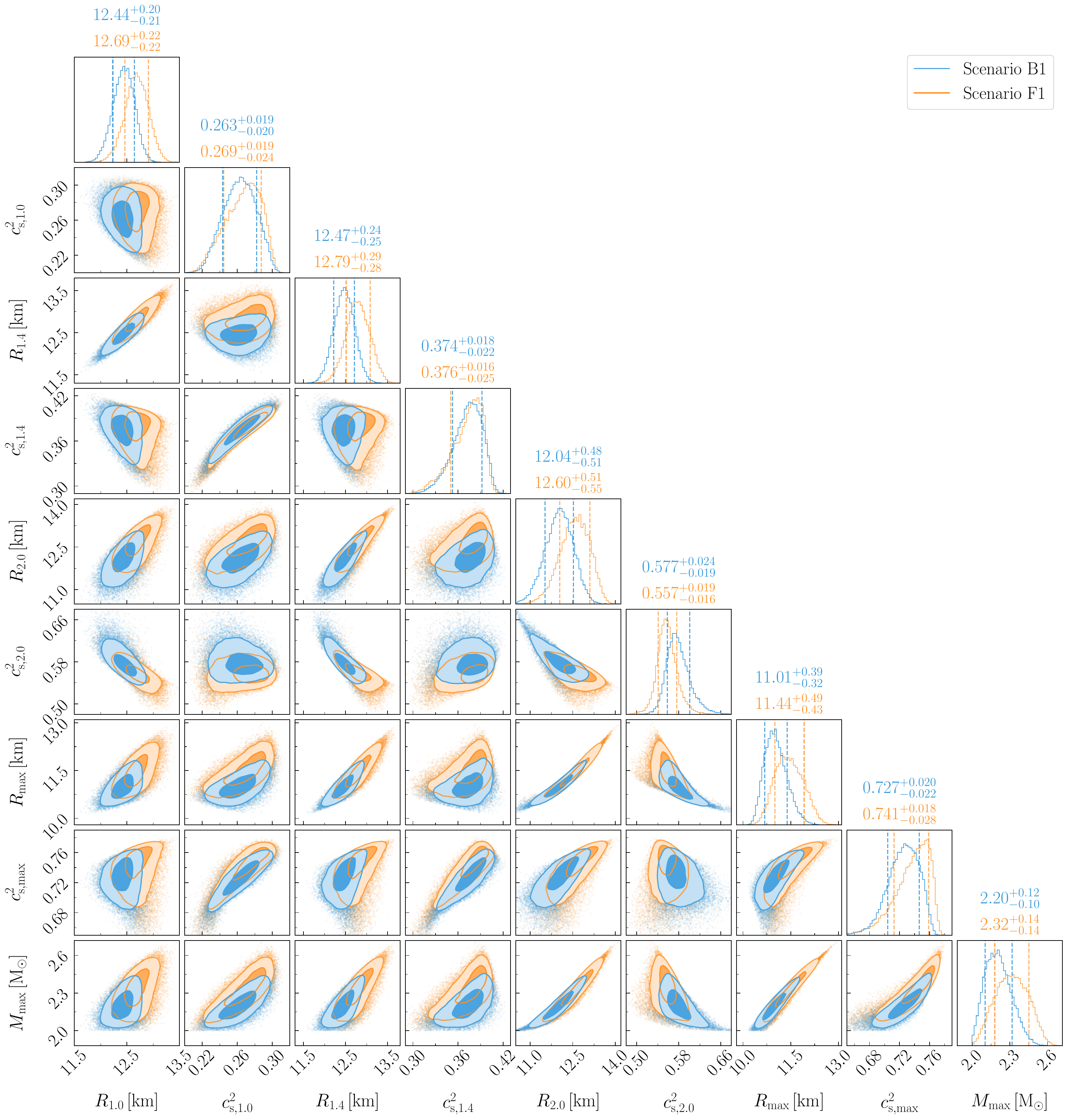}
\caption{
Posterior distributions of nine gross properties of compact stars:
radius $R$ and sound speed $c_{\rm s}^2$ at the center of the 
respective 1.0, 1.4, $2.0\,M_{\odot}$ and the maximum-mass 
$M_{\rm max}$ configurations, under astrophysical scenarios B1 
and F1. The light and dark shaded regions indicate respectively 
the 68.3\% and 95.4\% CIs of the two-dimensional distributions, 
while the one-dimensional posteriors for each quantity are 
given at the edges of  the plots where vertical lines indicate 
the 68.3\% CI. 
}
\label{fig:CS_properties}
\end{figure*}

\begin{figure*}[tb]
\centering
\includegraphics[width = 0.98\textwidth]{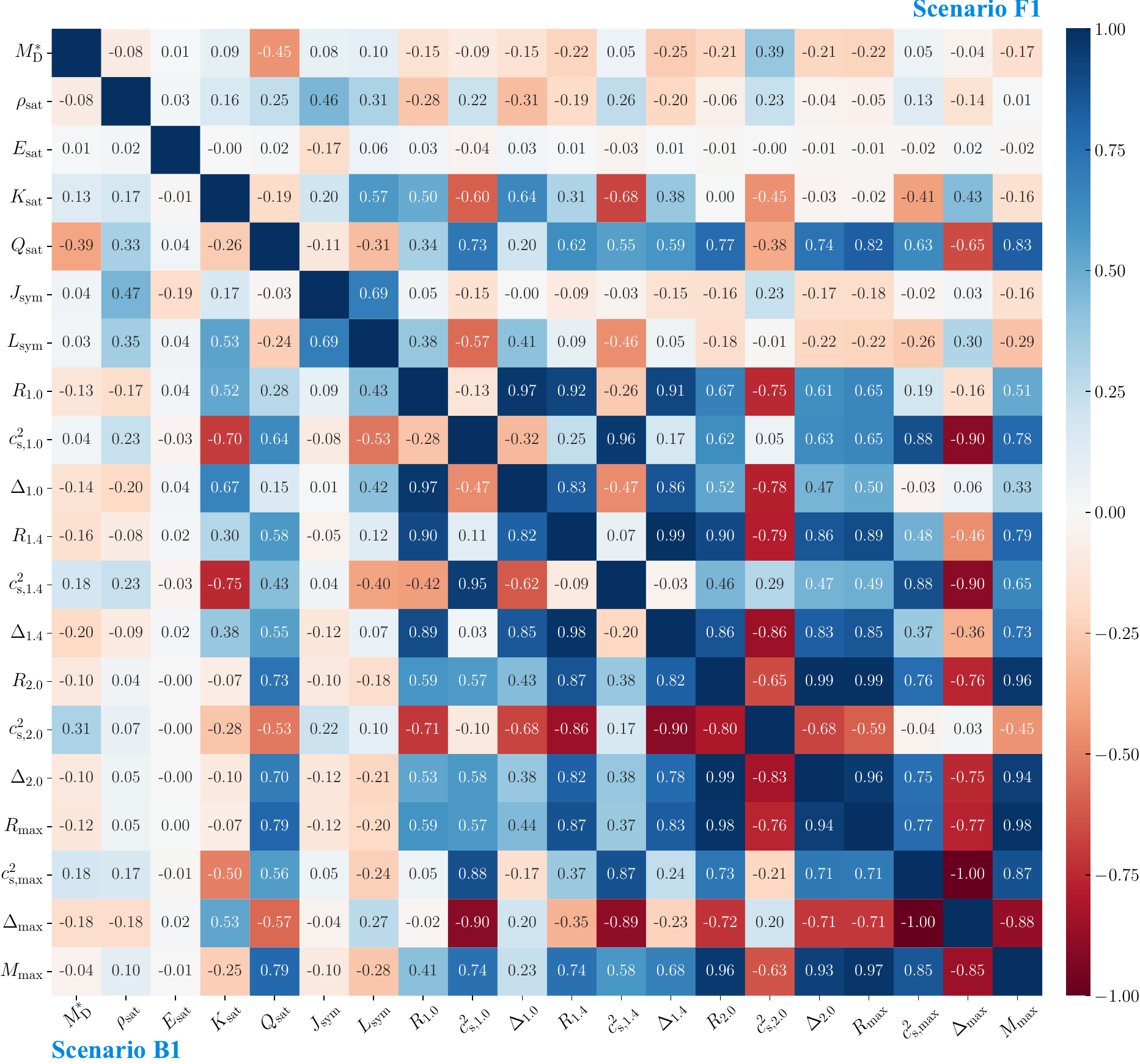}
\caption{
Posterior correlation matrix for variation of nuclear 
characteristic parameters at saturation density and 
selected gross properties of compact stars under astrophysical 
scenarios B1 (lower-left part) and F1 (upper-right part). 
}
\label{fig:CS_heatmap}
\end{figure*}

In this subsection, we assess the properties of the nucleonic EoS.
Figure~\ref{fig:EoS_scenarios} summarizes the posterior distributions 
for EoS used to generate the $M$-$R$ posteriors in the previous 
subsection, as well as the corresponding proton fraction under 
different astrophysical scenarios. For scenarios B1 and B4 
(F1 and F4), we show also the contours representing the positions of 
the respective $2.0\,M_{\odot}$ and the maximum-mass $M_{\rm max}$ 
configurations for reference. The differences for EoS posteriors 
among different scenarios are consistent with those for $M$-$R$
posteriors. 

For each scenario considered we find that the inclusion of HESS J1731
data produces slightly lower pressure for low-density regions
compared to posteriors for baseline. This is in agreement with its
low mass interval involved in modeling this object. The inclusion of
PSR J0952 data produces noticeably higher pressure at high-densities,
which is required to provide preference for values of
$M_{\rm max} \sim 2.3$-$2.4\,M_{\odot}$. As expected, low-density
$\rho\le 2\,\rho_{\rm sat}$ EoS posteriors are marginally affected 
by this additional constraint.

The impacts of the individual astrophysical constraints on EoS could be
understood in terms of the nuclear matter characteristics at
saturation density, which are shown in Fig.~\ref{fig:NM_characteristics}. 
For reference, we list here the selected characteristics predicted from 
the baseline scenarios B1 and F1. Scenario B1 features lower values for 
the incompressibility, $K_{\rm sat} = 231.4_{-33.3}^{+34.4}$~MeV and 
negative values for the skewness, 
$Q_{\rm sat} = -391.0_{-170.2}^{+333.6}$~MeV (at 68.3\% CI)
in the isoscalar sector, and symmetry energy
$J_{\rm sym} = 31.24_{-0.86}^{+0.88}$~MeV and its slope
$L_{\rm sym} = 46.73_{-5.67}^{+5.94}$~MeV in the isovector sector. The
scenario F1 has higher values for
$K_{\rm sat} = 244.8_{-34.7}^{+36.4}$~MeV and allows for positive
value for $Q_{\rm sat} = -164.7_{-280.1}^{+548.9}$~MeV (at 68.3\% CI)
in the isoscalar sector, and $J_{\rm sym} = 31.20_{-0.88}^{+0.89}$~MeV
and $L_{\rm sym} = 47.86_{-6.27}^{+6.46}$~MeV in the isovector sector.
Note that in this scenario the value of $Q_{\rm sat}$ can further
reach up to 900~MeV at a 95.4\% CI. The correlations among nuclear
characteristics at saturation density will be accessed below in
Subsec.~\ref{sec:Results_3}.

Including additional constraints, it is seen that the lowest-order
parameters $E_{\rm sat}$ and $J_{\rm sym}$ are almost identical among
scenarios B (F). The parameter $K_{\rm sat}$ is reduced by 5-10~MeV
($< 5\%$) and $L_{\rm sym}$ is reduced by only 1-2~MeV ($< 5\%$) among
scenarios B (F). These reductions are responsible for the softness
required by HESS J1731 data in scenarios B2 or F2. In contrast, for
scenarios B3 and F3 where the PSR J0952 data are included, such
reductions mainly countervail the stiffness induced by the higher
values of $Q_{\rm sat}$ and $K_{\rm sym}$ at low density regime, which
now are increased up to $Q_{\rm sat} = -215.0_{-253.0}^{+435.7}$~MeV
and $K_{\rm sym} = -96.8_{-16.6}^{+24.0}$~MeV  (at 68.3\% CI) for 
scenario B3, and to $Q_{\rm sat} = 81.0_{-386.6}^{+548.4}$~MeV and
$K_{\rm sym} = -83.8_{-19.5}^{+26.9}$~MeV for scenario F3. Note that
in our modeling, the parameters $K_{\rm sat}$ and $Q_{\rm sat}$ are
quasi-independent of each other, which is in contrast to previous
studies of this type CDF with simplified
functions~\eqref{eq:isoscalar_coupling} for density dependence of
isoscalar meson-nucleon
couplings~\citep{Malik:2022a,Beznogov:2023,Parmar:2024}. Since the 
isovector parameter $K_{\rm sym} $ is not entirely independent, 
its variations arise from slight adjustments in the two lower-order 
parameters, $J_{\rm sym}$ and $L_{\rm sym}$.

In the lower panels of Fig.~\ref{fig:EoS_scenarios} the horizontal
bands indicate the nucleonic direct Urca (DU) threshold. This band is,
however, model dependent~\citep{Klahn:2006}. For $\mu^-$ free case the
threshold value is 11.1\%; in the limit of massless $\mu$-s, which is
applicable for high densities matter, it yields an upper limit of
14.8\%. As seen in this plot, due to the softness (dominated by
$\chi$EFT constraint) of the nucleonic EoS at low densities, the DU
process will be mostly disallowed in CSs with $M \leq 2\,M_{\odot}$ if
one utilizes the current CDF-based EoS. 
The same conclusion was reached in 
Refs.~\citep{Malik:2022a,Beznogov:2024a,Char:2025}, 
where the density dependence of the $\rho$-meson coupling 
was also modeled using an exponential form.
However, note that the DU threshold may be smoothed out due to 
short-range correlations which allow for finite width of the baryon 
spectral function in dense matter~\citep{Sedrakian:2024}.
Moreover, non-nucleonic DU processes may still occur if hyperons or 
$\Delta$-resonances appear in the inner cores of compact stars; see, 
for example, Refs.~\citep{Lijj:2018a,Lijj:2018b,Fortin:2021}. Alternatively, 
modifications to the density dependence of the $\rho$-meson coupling 
function can lead to an earlier onset of nucleonic DU processes, as 
discussed in Refs.~\citep{Malik:2022b,Char:2023,Scurto:2024,Char:2025}.

Finally, in Fig.~\ref{fig:EoS_nuclear} we show the posterior 
distributions for the pressure (upper panels) and symmetry energy 
(lower panels) of symmetric nuclear matter as a function of baryonic 
density under different astrophysical scenarios. In each panel we 
also include for reference the experimental constraints obtained 
from extensive, independent studies.

In the upper panels of Fig.~\ref{fig:EoS_nuclear}, we show the
constraints for pressure obtained from heavy ion collision data. 
These include: (a) the constraint obtained by comparing measurements 
of collective flow from heavy-ion collisions with results from 
hadronic transport simulations using a set of EoSs (labeled
cFlow)~\citep{Danielewicz:2002}; (b) the constraint obtained by
reproducing in hadronic transport simulations with mean-field
interactions the ratios of experimentally measured kaon yields in
Au+Au and C+C collisions (Kanos)~\citep{Fuchs:2003,Lynch:2009}, and
(c) constraint obtained from the elliptic flow data measured by the
FOPI collaboration together with simulations from isospin quantum
molecular dynamics (QMD) (eFlow)~\citep{LeFevre:2016}. It is seen
that the 68.3\% CI regions of the EoSs from scenarios from B1-B4 
largely coincide with the HIC constraints, while the 95.4\% CI regions 
of the EoSs from scenarios F1-F4 overlap with the constraints. Note, 
however, that these HIC constraints exhibit some model dependence, 
especially regarding the high-density behavior explored in collision 
simulations. Moreover, certain nuclear models used to interpret the 
experimental data do not support two-solar-mass CSs.

As is well known, determining the composition at high densities
requires precise information on the density dependence of symmetry
energy. As shown in the lower panels of Fig.~\ref{fig:EoS_nuclear}, 
the predicted symmetry energy exhibits a remarkably narrow range. 
This is primarily due to the limited degrees of freedom in the isovector
sector of the present CDF model, which includes only two parameters,
both of which are strongly constrained by the $\chi$EFT results for
low-density pure neutron matter. Consequently, the behavior of
symmetry energy near $\rho_{\rm sat}$ and at higher densities is
predominantly governed by $L_{\rm sym}$. This, in turn, explains the
narrow band of proton fraction observed in the lower panels of
Fig.~\ref{fig:EoS_scenarios}.

For comparison, in the lower panels of Fig.~\ref{fig:EoS_nuclear} we
also give the constraints on symmetry energy obtained from: (a) the
comparison of the experimental measurements of isospin diffusion and
the ratio of neutron and proton spectra in collisions of Sn+Sn to
results from the improved QMD simulations (labeled
HIC)~\citep{Tsang:2009}; (b) the studies of the ratio of the elliptic
flow of neutrons and hydrogen nuclei in Au+Au collisions in the
FOPI-LAND experiment (FOPI)~\citep{Russotto:2011}; (c) the use of
neutron-to-charged fragments ratios measured in Au+Au collisions in
the ASY-EOS experiment (ASY)~\citep{Russotto:2016}; and (d) nuclear
structure studies involving excitation energies to isobaric analog
states (IAS)~\citep{Danielewicz:2014}.  Also shown in this plot is the
68.3\% CI region consistent with the best fit of experimental data on
the isospin diffusion in collision systems with a different
proton-to-neutron ratios, neutron-to-proton energy spectra, and
spectral pion ratios in Sn+Sn collisions (labeled FIT)~\citep{Lynch:2022}. 
It is clearly seen that our models are broadly consistent with these 
constraints. The predicted bulk quantities of nuclear matter at 
saturation density for different astrophysical scenarios are summarized 
in the Appendix.

\subsection{Correlations of CDF parameters, nuclear characteristics, 
and compact-star properties}
\label{sec:Results_3}
We now focus on the correlations of CDF parameters, nuclear matter
characteristics and CS properties, and the correlations
between two of them for representative scenarios. In
Fig.~\ref{fig:CDF_parameters}, we show the posterior distributions on
the CDF parameters for scenarios B1 and F1, respectively. For each,
we include the seven adjustable parameters
$(g_\sigma,\,g_\omega,\,g_\rho,\,a_\sigma,\,d_\sigma,\,d_\omega,\,a_\rho)$
that describe the CDF and in addition the parameter $a_\omega$. In these
corner plots, we show the two-dimensional (2D) correlated probability
distributions and along the diagonal one-dimensional (1D) probability
distribution functions (PDFs). The elliptical nature of the 2D
contour for a pair of parameters indicates the correlation existing
between them, while a circular nature indicates no correlation. In
the 1D plots, we also display the results that applied only the
constraints on the SNM characteristics at saturation density (labeled SNM) 
as well as those which combine these with the $\chi$EFT computations 
(labeled SNM+EFT).

As anticipated the pair of parameters $(g_\sigma,\,g_\omega)$ displayed
in Fig.~\ref{fig:CDF_parameters} high correlation, as these are the
parameters that determine the binding energy at saturation density. 
The pairs of $(d_\sigma,\,d_\omega)$ and $(a_\sigma,\,a_\omega)$ that 
control the density dependence of $\sigma$ and $\omega$ mesons' couplings
are also strongly correlated as they are constrained by the 
conditions~\eqref{eq:coupling_constraints}. In addition, the two 
parameters $(g_\rho,\,a_\rho)$ in the isovector sector were seen to 
be correlated already when only the $\chi$EFT constraint is applied, 
and their PDFs are almost identically to those for scenarios B1 and F1, 
which means that the astrophysical constraints do not introduce any 
noticeable modifications.

In Fig.~\ref{fig:NM_properties}, we show nine nuclear characteristic
parameters at saturation density, namely,
$(M^\ast_{\rm{D}},\,\rho_{\rm{sat}},\,E_{\rm{sat}},\,K_{\rm{sat}},\,Q_{\rm{sat}})$ 
in the isoscalar sector and $(J_{\rm{sym}},\,L_{\rm{sym}})$ in the
isovector sector, and for each sector we show an additional higher-order
parameter ($Z_{\rm sat}$ and $K_{\rm sym}$, respectively) which are
predictions based on values of the lower-order parameters that fix the
parameters of the present CDF. It is seen that the PDFs for parameters
$(M^\ast_{\rm{D}},\,\rho_{\rm{sat}},\,E_{\rm{sat}},\,K_{\rm{sat}})$ in
the isoscalar sector and $J_{\rm{sym}}$ in the isovector sector behave
similarly in different cases shown in Fig.~\ref{fig:NM_properties} as
the Gaussian distribution is applied to the priors. When the $\chi$EFT 
constraint is applied, the PDF for the slope parameter $L_{\rm{sym}}$ 
exhibits a Gaussian-like behavior. This arises from its strong 
correlation with $J_{\rm{sym}}$, despite the assumption of a uniform 
prior.

From Fig.~\ref{fig:NM_properties}, we observe that the lowest-order
parameters, $E_{\rm sat}$ and $J_{\rm sym}$, remain nearly
identical in both scenarios B1 and F1, indicating that astrophysical
constraints have no impact on these quantities. In contrast, the
higher-order parameters, $Q_{\rm sat}$ and $K_{\rm sym}$,
exhibit significant sensitivity to astrophysical constraints. However,
the variation in $K_{\rm sym }$ within the present CDF framework
arises primarily from slight rearrangements of the lower-order
isovector parameters $J_{\rm sym}$ and $L_{\rm sym}$.

Now, let us examine the correlations among CS properties.
Figure~\ref{fig:CS_properties} presents the posterior distributions
for nine key properties of CSs, including the radius $R$ and
the squared sound speed $c_{\rm{s}}^2=d P/d \varepsilon$ at the
center of stars with masses of $1.0,\,1.4$, and $2.0\,M_{\odot}$, as well
as for the maximum-mass $M_{\rm max}$ configuration, under scenarios
B1 and F1. The value of $c_{\rm{s}}^2$ characterizes the stiffness
of the EoS at the stellar center while $R$ reflects the overall
stiffness of the EoS from the core to the surface. As shown in
Fig.~\ref{fig:CS_properties}, the closer the masses of two stars,
the stronger the correlation between their radii or central sound
speeds. This trend is particularly evident for the present CDF-based
EoS models, which exhibit a relatively straightforward behavior.

Finally we present in Fig.~\ref{fig:CS_heatmap} the posterior
correlation matrix for variation of nuclear characteristic parameters
at saturation density and selected gross properties of CSs under
astrophysical scenarios B1 (in the lower-left) and F1 (in the
upper-right). In this figure, we introduce also the dimensionless
trace anomaly $\Delta = 1/3-P/\varepsilon$~\citep{Musolino:2024}.
The correlations among nuclear matter characteristics are larger in
their absolute values, compared to the case where only SNM constraints
are applied in Fig.~\ref{fig:NM_heatmap}. In particular, the
isovector pair $(J_{\rm sym}, L_{\rm sym})$ exhibits a
strong correlation, with a Pearson coefficient of approximately
$r \sim 0.7$. This correlation primarily arises from the $\chi$EFT
constraint on low-density PNM. The correlations among several key 
quantities of CSs are in agreement with our previous comments above.
Finally, we emphasize once more that in our analysis, the 
isoscalar pair $\left(K_{\rm{sat}}, Q_{\rm{sat}}\right)$ shows 
only a weak correlation ($r \sim 0.2$ ), whereas the pair 
$\left(K_{\rm{sat}}, Z_{\rm{sat }}\right)$ exhibits a strong correlation 
(see Fig.~\ref{fig:NM_properties}). This behavior is qualitatively 
consistent with the findings of Ref.~\citep{Providencia:2023}, where 
different observational constraints were employed. However, a recent 
study~\citep{Char:2025} reports a contrasting result: a strong correlation 
between $\left(K_{\rm{sat}}, Q_{\rm{sat}}\right)$ and an absence of 
correlation between $\left(K_{\rm{sat}}, Z_{\rm{sat}}\right)$. 
Further work is required to understand the sensitivity of correlations 
to the modeling framework and the choice of constraints.

\begingroup
\squeezetable
\begin{table*}[tb]
\caption{ Characteristic parameters of symmetric nuclear matter at
saturation density from the posterior distributions for scenarios
B. The last column (B$1_{\rm ne}$) corresponds to neglecting the 
$\chi$EFT constraint. The superscripts and subscripts indicate 
the 68.3\% and 95.4\% (in parentheses) CI ranges.
}
\setlength{\tabcolsep}{16.2pt}
\label{tab:NM_Posterior_B}
\centering
\begin{tabular}{ccccccc}
\hline \hline 
 Par.        & Unit        & 
 Scenario B1 & Scenario B2 & 
 Scenario B3 & Scenario B4 & 
 Scenario B1$_{\rm{ne}}$   \\      
\hline 
 $M_{\rm D}^\ast$            & $m_{\rm N}$     & 
 ${0.614}_{-0.047\,(0.098)}^{+0.039\,(0.072)}$ & 
 ${0.615}_{-0.047\,(0.095)}^{+0.039\,(0.071)}$ & 
 ${0.612}_{-0.043\,(0.091)}^{+0.034\,(0.063)}$ & 
 ${0.614}_{-0.043\,(0.090)}^{+0.033\,(0.062)}$ &
 ${0.613}_{-0.050\,(0.102)}^{+0.042\,(0.074)}$ \\
 $\rho_{\rm sat}$            & fm$^{-3}$       & 
 ${0.154}_{-0.005\,(0.009)}^{+0.005\,(0.009)}$ & 
 ${0.155}_{-0.005\,(0.009)}^{+0.005\,(0.010)}$ &
 ${0.155}_{-0.005\,(0.009)}^{+0.005\,(0.009)}$ & 
 ${0.155}_{-0.005\,(0.009)}^{+0.005\,(0.009)}$ & 
 ${0.154}_{-0.005\,(0.010)}^{+0.005\,(0.010)}$ \\
\hline
 $E_{\rm sat}$               & MeV           &
 ${-16.10}_{-0.20\,(0.40)}^{+0.20\,(0.40)}$  & 
 ${-16.11}_{-0.20\,(0.40)}^{+0.20\,(0.39)}$  &
 ${-16.11}_{-0.20\,(0.40)}^{+0.20\,(0.40)}$  & 
 ${-16.11}_{-0.20\,(0.40)}^{+0.20\,(0.39)}$  &     
 ${-16.09}_{-0.20\,(0.40)}^{+0.20\,(0.39)}$  \\
 $K_{\rm sat}$               & MeV           & 
 ${231.4}_{-33.3\,(64.0)}^{+34.4\,(68.7)}$   & 
 ${224.8}_{-32.5\,(63.8)}^{+34.2\,(69.7)}$   &
 ${225.4}_{-31.8\,(62.1)}^{+33.5\,(68.3)}$   & 
 ${219.7}_{-31.6\,(62.1)}^{+33.2\,(67.0)}$   &     
 ${232.9}_{-40.1\,(78.5)}^{+38.7\,(73.9)}$   \\
 $Q_{\rm sat}$               & MeV               & 
 ${-391.0}_{-170.2\,(323.3)}^{+333.6\,( 872.8)}$ & 
 ${-405.6}_{-171.1\,(312.9)}^{+315.8\,( 854.4)}$ & 
 ${-215.0}_{-253.0\,(437.5)}^{+435.7\,( 994.7)}$ & 
 ${-223.3}_{-258.6\,(438.0)}^{+421.6\,( 982.0)}$ &  
 ${-386.9}_{-181.5\,(410.0)}^{+369.1\,(1066.8)}$ \\
 $Z_{\rm sat}$               & MeV           & 
 ${4931}_{-2330\,(4721)}^{+1595\,(4862)}$    & 
 ${5174}_{-2170\,(4606)}^{+1477\,(4540)}$    &
 ${5393}_{-1836\,(4413)}^{+1483\,(5424)}$    & 
 ${5538}_{-1712\,(4183)}^{+1433\,(5073)}$    &     
 ${4944}_{-2730\,(5259)}^{+2178\,(9408)}$    \\
\hline  
 $J_{\rm sym}$               & MeV           & 
 ${31.24}_{-0.86\,(1.68)}^{+0.88\,(1.78)}$   & 
 ${31.20}_{-0.84\,(1.64)}^{+0.88\,(1.76)}$   &
 ${31.15}_{-0.84\,(1.64)}^{+0.89\,(1.79)}$   & 
 ${31.12}_{-0.83\,(1.64)}^{+0.88\,(1.75)}$   &     
 ${32.34}_{-2.02\,(4.01)}^{+1.96\,(3.96)}$   \\
 $L_{\rm sym}$               & MeV             &
 ${46.73}_{ -5.67\,(11.35)}^{ +5.94\,(12.10)}$ & 
 ${45.73}_{ -5.59\,(11.29)}^{ +5.64\,(11.79)}$ &
 ${45.46}_{ -5.79\,(11.20)}^{ +5.86\,(12.08)}$ & 
 ${44.45}_{ -5.65\,(11.21)}^{ +5.79\,(11.89)}$ & 
 ${45.05}_{-13.62\,(23.80)}^{+14.46\,(27.35)}$ \\
 $K_{\rm sym}$               & MeV           & 
 ${-106.0}_{-14.7\,(25.8)}^{+21.0\,(50.5)}$  & 
 ${-106.7}_{-14.7\,(25.5)}^{+21.2\,(52.0)}$  &
 ${ -96.8}_{-16.6\,(28.8)}^{+24.0\,(58.3)}$  & 
 ${ -96.9}_{-17.1\,(29.9)}^{+24.2\,(58.2)}$  &     
 ${-104.6}_{-25.7\,(48.5)}^{+35.3\,(83.9)}$  \\
 $Q_{\rm sym}$               & MeV             & 
 ${776.1}_{-159.1\,(323.0)}^{+155.5\,(295.8)}$ & 
 ${802.6}_{-151.1\,(312.5)}^{+149.4\,(288.2)}$ & 
 ${836.6}_{-155.6\,(318.7)}^{+147.0\,(265.3)}$ & 
 ${864.3}_{-152.9\,(317.0)}^{+139.8\,(256.3)}$ & 
 ${856.2}_{-381.7\,(682.4)}^{+368.7\,(676.0)}$ \\
 $Z_{\rm sym}$               & MeV           & 
 ${-5079}_{-1988\,(4606)}^{+1435\,(2523)}$   & 
 ${-5225}_{-1925\,(4585)}^{+1421\,(2496)}$   &
 ${-5979}_{-2205\,(5025)}^{+1664\,(2927)}$   & 
 ${-6155}_{-2177\,(4910)}^{+1680\,(2988)}$   & 
 ${-5965}_{-4714\,(9446)}^{+3376\,(4990)}$   \\
\hline\hline 
\end{tabular}
\end{table*}
\endgroup

Correlations between nuclear characteristics and CS 
properties are encoded in the Pearson coefficients $r$. We find, 
notably, that the previously reported correlation between the slope 
of the symmetry energy $L_{\rm sym}$ and the radius of a $1.4\,M_{\odot}$ 
star $R_{1.4}$~\cite{Horowitz:2001,Fattoyev:2013,Fortin:2016,Lijj:2019a}, 
appears to have largely vanished, with $r \sim 0.1$, as also observed 
in Refs.~\cite{Hornick:2018,Lijj:2024c}. However, a moderate
correlation remains between $L_{\rm sym}$ and the radius of a
$1.0\,M_{\odot}$ star - $R_{1.0}$, with a Pearson coefficient
of approximately $r \sim 0.4$. It is important to note that the
correlations reported in Refs.~\cite{Horowitz:2001,Lijj:2019a} were
observed in a set of models where the isoscalar sector was nearly
fixed and only the isovector parameters varied significantly.

\section{Conclusions and outlook}
\label{sec:Conclusions}
In this work, we extend the CDF-based Bayesian inference framework to
incorporate the latest multi-messenger constraints, including NICER's
$M$-$R$ estimates, gravitational wave events, and $\chi$EFT
computations. Additionally, we account for the heaviest known CS -- 
the ``black widow" pulsar PSR J0952-0607-- as well as the
$M$-$R$ estimates for the ultra-compact, low-mass object HESS
J1731-347. We adopt the class of CDF models with linear mesonic
terms and density-dependent meson-nucleon couplings, which have been
extensively tested against various nuclear phenomena, allowing us to
evaluate their flexibility and predictive power under more extreme
astrophysical conditions.

A systematic analysis is performed for typical astrophysical scenarios
where the importance of the individual impacts of PSR J0952-0607 and
HESS J1731-347 are elucidated by assessing the posterior distributions
for key properties of CSs, nuclear characteristics at saturation
density, and correlations among them, and the density dependence of
EoS and the symmetry energy.

We find that the present CDF models can naturally accommodate higher
maximum masses for CSs, reaching values of approximately
$2.4$-$2.5M_{\odot}$. This is accompanied by a noticeable increase in
the radii of massive CSs by about $0.2$-$0.4$ km. However, further 
softening of the low-density regime of the EoS to better match
the HESS J1731-347 data appears challenging, likely requiring more
refined modeling of the star itself, including phase transitions to
quark matter, or further extensions of the CDF framework.
Nevertheless, we imposed tighter limits on the parameter space of
present CDF models which appear to be broadly consistent with current
nuclear experimental and astrophysical data. We find that the two
isovector parameters are tightly constrained at low-densities close to
saturation, leaving little room for variation in the high-density
symmetry energy. As a result, the symmetry energy predicted by the CDF
model is soft.  Furthermore, the softness of the EoS at low densities 
suppresses the nucleonic direct Urca (DU) cooling process in CSs with 
$M \leq 2 M_{\odot}$, as dictated by the behavior of the symmetry energy 
in the CDF model. A summary of the key predicted quantities for nuclear 
matte and CSs across various astrophysical scenarios is provided in the
Appendix.

In conclusion, this work underscores the importance of combining
multimessenger observations and advanced theoretical frameworks to
enhance the fidelity of nuclear astrophysical models of CSs. The CDF
models with density-dependent meson-nucleon couplings provide a robust
theoretical framework to interpret currently known gross properties of
CS. Due to the limited degrees of freedom in the isovector sector of 
the current CDF formulation, the predicted lower limit for the radii of 
low-mass stars cannot be reduced much below 12~km. Consequently, 
refining and extending the density dependence of CDF parameters is 
necessary to accurately model potential ultra-compact low-mass stars 
within the nucleonic paradigm. (Recall that a phase transition to 
quark matter may be an alternative solution as indicated above, see 
Refs.~\cite{Lijj:2024a,Lijj:2024b} and references therein). Additional 
parameter(s) could also be useful within the CDF framework to provide 
greater control over the symmetry energy’s high-density behavior, 
particularly through the parameter $K_{\rm sym}$. This issue will be 
further investigated in an upcoming study.

\section*{Acknowledgments}
J.~J.~L. and Y.~T. acknowledge the support of the National Natural 
Science Foundation of China under Grants No. 12105232 and No. 12475150. 
A.~S. acknowledges funding by the Deutsche Forschungsgemeinschaft Grant 
No. SE 1836/6-1 and the Polish NCN Grant No. 2023/51/B/ST9/02798.

\appendix*
\section{Key quantities of nuclear matter and compact stars}
\label{appendix}

\begingroup
\squeezetable
\begin{table*}[tb]
\caption{ Key quantities of compact stars from the posterior
distributions for scenarios B: radii, TDs, central baryonic
densities, pressures, energy densities, sound speeds, and trace
anomalies for $1.0,\,1.4,\,2.0\,M_{\odot}$ and the maximum-mass
$M_{\rm max}$ stars. We also list the inferred radius differences 
between $1.4,\,2.0\,M_{\odot}$ and $1.0,\,1.4\,M_{\odot}$ stars.  
The last column (B$1_{\rm ne}$) corresponds to neglecting the 
$\chi$EFT constraint. The superscripts and subscripts indicate 
the 68.3\% and 95.4\% (in parentheses) CI ranges.}
\setlength{\tabcolsep}{12.4pt}
\label{tab:CS_Posterior_B}
\centering
\begin{tabular}{ccccccc}
\hline \hline 
Par.                    & Unit             & 
Scenario B1             & Scenario B2      &   
Scenario B3             & Scenario B4      &  
Scenario B1$_{\rm{ne}}$                    \\  
\hline                          
    $R_{1.0}$                  &  km       &
${12.44}_{-0.21\,(0.43)}^{+0.20\,(0.38)}$  &  
${12.36}_{-0.22\,(0.42)}^{+0.21\,(0.40)}$  &   
${12.51}_{-0.20\,(0.41)}^{+0.18\,(0.37)}$  &  
${12.44}_{-0.21\,(0.41)}^{+0.20\,(0.38)}$  &    
${12.46}_{-0.27\,(0.51)}^{+0.27\,(0.56)}$  \\
    $\Lambda_{1.0}$            &           & 
${3114}_{-323\,(619)}^{+331\,(667)}$       &  
${2999}_{-319\,(596)}^{+346\,(683)}$       & 
${3259}_{-317\,(618)}^{+321\,(675)}$       &  
${3159}_{-320\,(614)}^{+341\,(683)}$       &    
${3176}_{-341\,(653)}^{+363\,(776)}$      \\
    $\rho_{1.0}$               &  fm$^{-3}$   &
${0.342}_{-0.021\,(0.041)}^{+0.022\,(0.044)}$ &  
${0.348}_{-0.022\,(0.042)}^{+0.023\,(0.045)}$ &  
${0.329}_{-0.019\,(0.037)}^{+0.020\,(0.042)}$ &  
${0.333}_{-0.020\,(0.039)}^{+0.021\,(0.044)}$ &    
${0.338}_{-0.021\,(0.042)}^{+0.022\,(0.046)}$ \\
    $P_{1.0}$                  &  MeV/fm$^{3}$ & 
${30.27}_{-2.37\,(4.52)}^{+2.69\,(5.69)}$      &  
${31.07}_{-2.56\,(4.81)}^{+2.86\,(5.84)}$      &
${28.94}_{-2.15\,(4.17)}^{+2.45\,(5.17)}$      &  
${29.56}_{-2.31\,(4.41)}^{+2.62\,(5.51)}$      &    
${29.79}_{-2.33\,(4.58)}^{+2.65\,(5.71)}$      \\
    $\varepsilon_{1.0}$        &  MeV/fm$^{3}$ & 
${340.87}_{-21.65\,(42.40)}^{+22.58\,(45.95)}$ &  
${346.57}_{-22.41\,(44.03)}^{+23.87\,(47.02)}$ &   
${327.08}_{-19.66\,(38.61)}^{+21.10\,(43.37)}$ &  
${331.63}_{-20.49\,(40.26)}^{+22.21\,(45.91)}$ &    
${336.81}_{-22.10\,(43.96)}^{+23.06\,(48.06)}$ \\
    $c_{s, 1.0}^2$             &               &
${0.263}_{-0.020\,(0.037)}^{+0.019\,(0.033)}$  &  
${0.265}_{-0.019\,(0.036)}^{+0.018\,(0.032)}$  &  
${0.274}_{-0.018\,(0.037)}^{+0.016\,(0.026)}$  &  
${0.276}_{-0.018\,(0.037)}^{+0.015\,(0.025)}$  &   
${0.264}_{-0.026\,(0.047)}^{+0.027\,(0.047)}$  \\
    $\Delta_{1.0}$             &               & 
${0.244}_{-0.002\,(0.005)}^{+0.002\,(0.004)}$  &  
${0.244}_{-0.002\,(0.005)}^{+0.002\,(0.004)}$  &   
${0.245}_{-0.002\,(0.004)}^{+0.002\,(0.004)}$  &  
${0.244}_{-0.002\,(0.004)}^{+0.002\,(0.004)}$  &  
${0.245}_{-0.003\,(0.005)}^{+0.003\,(0.006)}$  \\ 
\hline     
    $R_{1.4}$                  &  km        &
${12.47}_{-0.25\,(0.50)}^{+0.24\,(0.48)}$   &  
${12.39}_{-0.26\,(0.50)}^{+0.26\,(0.49)}$   &
${12.61}_{-0.24\,(0.48)}^{+0.23\,(0.46)}$   &  
${12.54}_{-0.25\,(0.50)}^{+0.24\,(0.47)}$   & 
${12.50}_{-0.25\,(0.51)}^{+0.25\,(0.50)}$   \\  
    $\Lambda_{1.4}$            &            & 
${472}_{-65\,(121)}^{+76\,(163)}$           &  
${454}_{-65\,(118)}^{+75\,(161)}$           &  
${519}_{-68\,(129)}^{+76\,(163)}$           &  
${503}_{-69\,(130)}^{+76\,(164)}$           &  
${486}_{-68\,(129)}^{+78\,(174)}$          \\ 
    $\rho_{1.4}$               &  fm$^{-3}$   &
${0.425}_{-0.035\,(0.067)}^{+0.036\,(0.070)}$ &  
${0.432}_{-0.036\,(0.070)}^{+0.036\,(0.071)}$ &  
${0.400}_{-0.030\,(0.057)}^{+0.032\,(0.067)}$ &  
${0.406}_{-0.031\,(0.059)}^{+0.034\,(0.070)}$ & 
${0.419}_{-0.037\,(0.071)}^{+0.038\,(0.075)}$ \\  
    $P_{1.4}$                  & MeV/fm$^{3}$    & 
${60.042}_{-6.823\,(12.859)}^{+7.485\,(15.313)}$ & 
${61.763}_{-7.160\,(13.520)}^{+7.808\,(15.820)}$ & 
${55.432}_{-5.778\,(10.937)}^{+6.564\,(13.889)}$ & 
${56.661}_{-6.052\,(11.422)}^{+7.011\,(14.940)}$ & 
${58.872}_{-7.017\,(13.282)}^{+7.696\,(16.131)}$ \\ 
     $\varepsilon_{1.4}$        & MeV/fm$^{3}$   & 
${433.76}_{-38.65\,(73.38)}^{+39.20\,(76.56)}$   & 
${441.75}_{-39.51\,(75.61)}^{+39.89\,(77.99)}$   &  
${406.84}_{-32.54\,(61.79)}^{+35.26\,(72.82)}$   & 
${412.46}_{-33.66\,(63.60)}^{+37.27\,(76.79)}$   & 
${427.67}_{-41.19\,(77.86)}^{+42.64\,(83.86)}$   \\
     $c_{s, 1.4}^2$             &                & 
${0.374}_{-0.022\,(0.046)}^{+0.018\,(0.030)}$    & 
${0.377}_{-0.020\,(0.044)}^{+0.017\,(0.029)}$    & 
${0.383}_{-0.019\,(0.041)}^{+0.014\,(0.023)}$    & 
${0.385}_{-0.018\,(0.040)}^{+0.013\,(0.022)}$    &   
${0.375}_{-0.027\,(0.051)}^{+0.023\,(0.038)}$    \\ 
     $\Delta_{1.4}$             &                & 
${0.195}_{-0.005\,(0.010)}^{+0.004\,(0.008)}$    & 
${0.193}_{-0.005\,(0.010)}^{+0.005\,(0.009)}$    & 
${0.197}_{-0.004\,(0.009)}^{+0.004\,(0.008)}$    & 
${0.196}_{-0.005\,(0.010)}^{+0.004\,(0.008)}$    &  
${0.196}_{-0.005\,(0.010)}^{+0.004\,(0.008)}$    \\
\hline     
    $R_{2.0}$                  & km            &
${12.04}_{-0.51\,(1.03)}^{+0.48\,(0.91)}$      & 
${11.95}_{-0.52\,(1.02)}^{+0.49\,(0.93)}$      &  
${12.38}_{-0.44\,(0.92)}^{+0.40\,(0.77)}$      & 
${12.31}_{-0.45\,(0.96)}^{+0.41\,(0.79)}$      &  
${12.11}_{-0.55\,(1.11)}^{+0.49\,(0.95)}$      \\
    $\Lambda_{2.0}$            &               &
${33.2}_{-11.1\,(19.1)}^{+14.2\,(30.6)}$       & 
${31.1}_{-10.5\,(17.7)}^{+13.6\,(29.7)}$       &  
${42.8}_{-12.0\,(21.9)}^{+13.9\,(29.7)}$       & 
${40.9}_{-11.9\,(21.7)}^{+13.7\,(29.3)}$       &  
${35.0}_{-12.7\,(21.2)}^{+16.1\,(34.9)}$       \\ 
    $\rho_{2.0}$               & fm$^{-3}$     &
${0.636}_{-0.093\,(0.162)}^{+0.122\,(0.273)}$  &  
${0.653}_{-0.097\,(0.169)}^{+0.123\,(0.266)}$  & 
${0.568}_{-0.067\,(0.120)}^{+0.088\,(0.205)}$  & 
${0.578}_{-0.070\,(0.123)}^{+0.093\,(0.219)}$  &  
${0.623}_{-0.099\,(0.169)}^{+0.134\,(0.291)}$  \\
    $P_{2.0}$                  & MeV/fm$^{3}$   &
${193.66}_{-46.59\,(76.50)}^{+72.30\,(185.95)}$ & 
${202.93}_{-49.51\,(81.60)}^{+76.65\,(188.17)}$ &
${158.85}_{-30.66\,(52.38)}^{+45.43\,(118.10)}$ & 
${164.06}_{-32.46\,(55.09)}^{+49.27\,(130.99)}$ &  
${186.18}_{-47.67\,(77.15)}^{+79.54\,(199.82)}$ \\ 
    $\varepsilon_{2.0}$        & MeV/fm$^{3}$      &
${704.78}_{-118.38\,(201.48)}^{+164.86\,(385.85)}$ & 
${726.13}_{-123.53\,(211.30)}^{+167.48\,(379.18)}$ & 
${617.12}_{-82.44\,(145.17)}^{+112.83\,(273.48)}$  & 
${629.51}_{-85.86\,(149.94)}^{+119.41\,(293.40)}$  &  
${687.99}_{-125.32\,(208.97)}^{+181.32\,(411.07)}$ \\
    $c_{s, 2.0}^2$             &                & 
${0.577}_{-0.019\,(0.036)}^{+0.024\,(0.057)}$   & 
${0.582}_{-0.020\,(0.038)}^{+0.025\,(0.059)}$   &  
${0.569}_{-0.015\,(0.030)}^{+0.018\,(0.040)}$   & 
${0.572}_{-0.016\,(0.031)}^{+0.019\,(0.044)}$   &  
${0.575}_{-0.021\,(0.043)}^{+0.026\,(0.064)}$   \\
    $\Delta_{2.0}$             &                & 
${0.059}_{-0.032\,(0.074)}^{+0.024\,(0.042)}$   & 
${0.054}_{-0.033\,(0.075)}^{+0.025\,(0.044)}$   & 
${0.076}_{-0.023\,(0.054)}^{+0.018\,(0.032)}$   & 
${0.073}_{-0.025\,(0.059)}^{+0.019\,(0.033)}$   &   
${0.063}_{-0.035\,(0.082)}^{+0.025\,(0.043)}$   \\ 
\hline     
    $M_{\rm max}$              & $M_{\odot}$   & 
${2.20}_{-0.10\,(0.17)}^{+0.12\,(0.23)}$       & 
${2.19}_{-0.10\,(0.16)}^{+0.11\,(0.23)}$       &
${2.28}_{-0.10\,(0.19)}^{+0.10\,(0.20)}$       & 
${2.27}_{-0.10\,(0.19)}^{+0.10\,(0.20)}$       &  
${2.21}_{-0.11\,(0.19)}^{+0.13\,(0.26)}$       \\
     $R_{M_{\rm max}}$          & km           &
${11.00}_{-0.32\,(0.58)}^{+0.39\,(0.80)}$      & 
${10.94}_{-0.32\,(0.57)}^{+0.38\,(0.79)}$      &  
${11.27}_{-0.34\,(0.63)}^{+0.36\,(0.73)}$      & 
${11.22}_{-0.34\,(0.64)}^{+0.36\,(0.74)}$      & 
${11.05}_{-0.35\,(0.62)}^{+0.42\,(0.88)}$      \\
     $\Lambda_{M_{\rm max}}$    &              &
${6.04}_{-0.58\,(0.93)}^{+0.78\,(1.78)}$       & 
${6.05}_{-0.56\,(0.91)}^{+0.69\,(1.59)}$       & 
${5.64}_{-0.40\,(0.64)}^{+0.61\,(1.45)}$       & 
${5.64}_{-0.39\,(0.61)}^{+0.57\,(1.36)}$       &  
${6.02}_{-0.68\,(1.02)}^{+0.91\,(2.05)}$      \\
    $\rho_{M_{\rm max}}$       & fm$^{-3}$     & 
${0.996}_{-0.079\,(0.152)}^{+0.075\,(0.136)}$  & 
${1.008}_{-0.079\,(0.153)}^{+0.073\,(0.132)}$  & 
${0.939}_{-0.066\,(0.127)}^{+0.070\,(0.139)}$  & 
${0.948}_{-0.068\,(0.128)}^{+0.072\,(0.142)}$  & 
${0.986}_{-0.088\,(0.166)}^{+0.083\,(0.148)}$  \\
    $P_{M_{\rm max}}$          & MeV/fm$^{3}$  &
${588.95}_{-31.28\,(62.70)}^{+30.94\,(62.09)}$ & 
${596.25}_{-31.84\,(64.58)}^{+32.32\,(62.72)}$ &  
${569.79}_{-29.47\,(58.40)}^{+29.69\,(60.19)}$ & 
${575.54}_{-30.51\,(60.15)}^{+30.96\,(62.88)}$ & 
${582.77}_{-35.70\,(68.17)}^{+36.75\,(74.48)}$ \\ 
    $\varepsilon_{M_{\rm max}}$ &  MeV/fm$^{3}$     & 
${1298.90}_{-104.00\,(199.98)}^{+97.16\,(177.49)}$  & 
${1315.11}_{-103.93\,(201.79)}^{+95.49\,(174.06)}$  &
${1224.77}_{ -87.64\,(167.09)}^{+92.29\,(181.76)}$  & 
${1236.73}_{ -89.22\,(169.07)}^{+94.77\,(186.52)}$  &  
${1285.11}_{-115.70\,(218.44)}^{+110.22\,(198.89)}$ \\
    $c_{s, M_{\rm max}}^2$     &               & 
${0.726}_{-0.023\,(0.049)}^{+0.020\,(0.032)}$  & 
${0.726}_{-0.020\,(0.044)}^{+0.019\,(0.032)}$  & 
${0.740}_{-0.019\,(0.043)}^{+0.014\,(0.023)}$  & 
${0.739}_{-0.018\,(0.040)}^{+0.014\,(0.023)}$  & 
${0.728}_{-0.026\,(0.057)}^{+0.022\,(0.034)}$  \\ 
    $\Delta_{M_{\rm max}}$     &               & 
${-0.123}_{-0.016\,(0.027)}^{+0.018\,(0.038)}$ & 
${-0.123}_{-0.015\,(0.026)}^{+0.016\,(0.035)}$ &
${-0.134}_{-0.012\,(0.019)}^{+0.016\,(0.035)}$ & 
${-0.134}_{-0.011\,(0.019)}^{+0.015\,(0.033)}$ &  
${-0.124}_{-0.019\,(0.029)}^{+0.021\,(0.044)}$ \\  
\hline                                            
   $R_{1.4}$-$R_{2.0}$        & km         & 
${0.41}_{-0.27\,(0.46)}^{+0.34\,(0.74)}$   & 
${0.43}_{-0.27\,(0.47)}^{+0.33\,(0.71)}$   & 
${0.21}_{-0.19\,(0.34)}^{+0.25\,(0.58)}$   & 
${0.22}_{-0.19\,(0.34)}^{+0.26\,(0.59)}$   &
${0.37}_{-0.33\,(0.54)}^{+0.42\,(0.87)}$   \\
    $R_{1.4}$-$R_{1.0}$        & km        & 
${0.03}_{-0.11\,(0.20)}^{+0.11\,(0.21)}$   & 
${0.03}_{-0.10\,(0.20)}^{+0.11\,(0.21)}$   &
${0.11}_{-0.10\,(0.20)}^{+0.10\,(0.18)}$   & 
${0.10}_{-0.10\,(0.20)}^{+0.09\,(0.18)}$   & 
${0.05}_{-0.19\,(0.37)}^{+0.16\,(0.28)}$   \\ 
\hline\hline
\end{tabular}
\end{table*}
\endgroup

\begingroup
\squeezetable
\begin{table*}[tb]
\caption{
The same as Table~\ref{tab:NM_Posterior_B}, but for scenarios F.}
\setlength{\tabcolsep}{16.4pt}
\label{tab:NM_Posterior_F}
\centering
\begin{tabular}{ccccccc}
\hline \hline 
 Par.        & Unit        &       
 Scenario F1 & Scenario F2 &
 Scenario F3 & Scenario F4 & 
 Scenario F1$_{\rm{ne}}$   \\
\hline 
 $M_{\rm D}^\ast$            & $m_{\rm N}$     & 
 ${0.605}_{-0.043\,(0.091)}^{+0.035\,(0.070)}$ &
 ${0.606}_{-0.043\,(0.092)}^{+0.036\,(0.070)}$ &
 ${0.601}_{-0.039\,(0.086)}^{+0.030\,(0.059)}$ & 
 ${0.603}_{-0.039\,(0.086)}^{+0.030\,(0.060)}$ & 
 ${0.607}_{-0.047\,(0.098)}^{+0.041\,(0.075)}$ \\ 
 $\rho_{\rm sat}$            & fm$^{-3}$       & 
 ${0.154}_{-0.005\,(0.009)}^{+0.005\,(0.009)}$ & 
 ${0.154}_{-0.005\,(0.009)}^{+0.005\,(0.009)}$ &
 ${0.154}_{-0.005\,(0.009)}^{+0.005\,(0.009)}$ & 
 ${0.154}_{-0.004\,(0.009)}^{+0.005\,(0.009)}$ & 
 ${0.153}_{-0.005\,(0.010)}^{+0.005\,(0.010)}$ \\ 
\hline
 $E_{\rm sat}$               & MeV           &
 ${-16.10}_{-0.20\,(0.40)}^{+0.20\,(0.40)}$  & 
 ${-16.10}_{-0.20\,(0.40)}^{+0.20\,(0.40)}$  & 
 ${-16.11}_{-0.19\,(0.39)}^{+0.20\,(0.40)}$  & 
 ${-16.11}_{-0.20\,(0.40)}^{+0.20\,(0.40)}$  & 
 ${-16.09}_{-0.20\,(0.40)}^{+0.20\,(0.40)}$  \\ 
 $K_{\rm sat}$               & MeV           &  
 ${244.8}_{-34.7\,(67.8)}^{+36.4\,(71.2)}$   & 
 ${236.9}_{-34.1\,(66.4)}^{+36.8\,(73.6)}$   & 
 ${241.2}_{-32.7\,(63.8)}^{+33.7\,(70.3)}$   & 
 ${234.0}_{-32.4\,(63.3)}^{+33.9\,(69.9)}$   & 
 ${240.2}_{-40.1\,(79.8)}^{+40.9\,(78.0)}$   \\
 $Q_{\rm sat}$               & MeV              & 
 ${-164.7}_{-280.1\,(455.4)}^{+548.9\,(1161.9)}$& 
 ${-195.8}_{-271.8\,(444.7)}^{+544.8\,(1163.0)}$&  
 ${81.0}_{-386.6\,(625.9)}^{+548.4\,(1071.4)}$  & 
 ${47.1}_{-375.9\,(604.3)}^{+540.6\,(1069.8)}$  & 
 ${-285.5}_{-244.6\,(476.7)}^{+560.4\,(1340.1)}$\\
 $Z_{\rm sat}$               & MeV            & 
 ${4641}_{-2508\,(5071)}^{+1663\,( 5497)}$    & 
 ${4872}_{-2314\,(5105)}^{+1623\,( 5381)}$    & 
 ${4877}_{-1928\,(4831)}^{+1796\,( 5729)}$    & 
 ${5053}_{-1957\,(4790)}^{+1674\,( 5437)}$    & 
 ${4707}_{-2916\,(5468)}^{+2452\,(11286)}$    \\
\hline         
 $J_{\rm sym}$               & MeV           & 
 ${31.20}_{-0.88\,(1.72)}^{+0.89\,(1.82)}$   & 
 ${31.15}_{-0.85\,(1.69)}^{+0.91\,(1.83)}$   & 
 ${31.12}_{-0.86\,(1.70)}^{+0.88\,(1.79)}$   & 
 ${31.05}_{-0.84\,(1.65)}^{+0.88\,(1.78)}$   & 
 ${32.25}_{-2.03\,(3.99)}^{+2.03\,(4.00)}$   \\
 $L_{\rm sym}$               & MeV             &
 ${47.86}_{ -6.27\,(12.15)}^{ +6.46\,(13.08)}$ & 
 ${46.62}_{ -6.14\,(11.97)}^{ +6.36\,(13.00)}$ & 
 ${46.75}_{ -6.12\,(11.87)}^{ +6.38\,(13.15)}$ & 
 ${45.46}_{ -6.02\,(11.62)}^{ +6.22\,(12.77)}$ & 
 ${64.03}_{-21.16\,(37.18)}^{+16.66\,(28.09)}$ \\
 $K_{\rm sym}$               & MeV           & 
 ${-94.2}_{-18.2\,(31.1)}^{+26.6\,(61.8)}$   & 
 ${-95.1}_{-18.3\,(31.6)}^{+27.5\,(62.3)}$   & 
 ${-83.8}_{-19.5\,(34.1)}^{+26.9\,(60.7)}$   & 
 ${-84.1}_{-20.1\,(35.0)}^{+27.9\,(62.5)}$   & 
 ${-83.3}_{-30.4\,(55.5)}^{+39.4\,(78.9)}$   \\
 $Q_{\rm sym}$               & MeV             & 
 ${765.3}_{-177.0\,(358.4)}^{+166.5\,(299.3)}$ & 
 ${796.9}_{-174.7\,(354.3)}^{+161.8\,(290.8)}$ & 
 ${817.2}_{-166.5\,(347.3)}^{+145.1\,(263.3)}$ & 
 ${847.2}_{-163.1\,(340.3)}^{+141.3\,(257.3)}$ & 
 ${389.6}_{-339.0\,(565.2)}^{+550.8\,(997.8)}$ \\ 
 $Z_{\rm sym}$               & MeV           & 
 ${-5686}_{-2596\,(5565)}^{+1820\,(3095)}$   & 
 ${-5853}_{-2566\,(5491)}^{+1835\,(3119)}$   & 
 ${-6702}_{-2563\,(5287)}^{+2024\,(3508)}$   & 
 ${-6857}_{-2577\,(5323)}^{+2002\,(3538)}$   & 
 ${-2740}_{-5076\,(11484)}^{+1891\,(2808)}$  \\
\hline \hline 
\end{tabular}
\end{table*}
\endgroup

\begingroup
\squeezetable
\begin{table*}[tb]
\caption{
The same as Table~\ref{tab:CS_Posterior_B}, but for scenarios F.}
\setlength{\tabcolsep}{12.4pt}
\label{tab:CS_Posterior_F}
\centering
\begin{tabular}{ccccccc}
\hline \hline 
 Par.                       & Unit          &  
 Scenario F1                & Scenario F2   & 
 Scenario F3                & Scenario F4   &         
 Scenario F1$_{\rm{ne}}$                    \\
    \hline                          
 $R_{1.0}$                  &  km            &
 ${12.69}_{-0.22\,(0.47)}^{+0.22\,(0.43)}$   &  
 ${12.61}_{-0.25\,(0.50)}^{+0.22\,(0.43)}$   & 
 ${12.75}_{-0.21\,(0.44)}^{+0.21\,(0.41)}$   &  
 ${12.68}_{-0.22\,(0.45)}^{+0.21\,(0.40)}$   &      
 ${13.05}_{-0.44\,(0.81)}^{+0.44\,(0.75)}$   \\
 $\Lambda_{1.0}$            &               & 
 ${3554}_{-388\,(764)}^{+429\,(858)}$       &  
 ${3429}_{-412\,(790)}^{+405\,(822)}$       &
 ${3712}_{-383\,(740)}^{+413\,(799)}$       &  
 ${3584}_{-376\,(736)}^{+401\,(773)}$       &      
 ${4046}_{-622\,(1104)}^{+756\,(1465)}$     \\
 $\rho_{1.0}$               &  fm$^{-3}$       &
 ${0.316}_{-0.022\,(0.040)}^{+0.023\,(0.047)}$ &   
 ${0.322}_{-0.023\,(0.042)}^{+0.024\,(0.051)}$ &   
 ${0.304}_{-0.019\,(0.033)}^{+0.021\,(0.043)}$ & 
 ${0.310}_{-0.019\,(0.033)}^{+0.022\,(0.044)}$ &      
 ${0.308}_{-0.021\,(0.036)}^{+0.023\,(0.047)}$ \\
 $P_{1.0}$                  &  MeV/fm$^{3}$   & 
 ${27.19}_{-2.46\,(4.43)}^{+2.66\,(5.69)}$    & 
 ${27.94}_{-2.52\,(4.61)}^{+2.93\,(6.31)}$    &  
 ${26.06}_{-2.13\,(3.77)}^{+2.41\,(5.03)}$    & 
 ${26.70}_{-2.14\,(3.80)}^{+2.51\,(5.31)}$    &      
 ${25.46}_{-2.54\,(4.46)}^{+2.99\,(6.11)}$   \\
 $\varepsilon_{1.0}$        &  MeV/fm$^{3}$     & 
 ${313.91}_{-23.39\,(41.88)}^{+23.82\,(48.94)}$ & 
 ${319.79}_{-23.93\,(43.46)}^{+25.06\,(52.72)}$ &   
 ${301.73}_{-19.38\,(34.30)}^{+21.78\,(44.24)}$ & 
 ${307.07}_{-19.64\,(34.55)}^{+22.21\,(45.92)}$ &      
 ${306.92}_{-21.29\,(37.60)}^{+23.90\,(48.44)}$ \\
 $c_{s, 1.0}^2$             &                   & 
 ${0.269}_{-0.024\,(0.044)}^{+0.019\,(0.030)}$  & 
 ${0.271}_{-0.023\,(0.044)}^{+0.018\,(0.029)}$  &  
 ${0.278}_{-0.019\,(0.041)}^{+0.014\,(0.023)}$  &  
 ${0.279}_{-0.018\,(0.040)}^{+0.013\,(0.022)}$  & 
 ${0.242}_{-0.037\,(0.060)}^{+0.039\,(0.065)}$  \\
 $\Delta_{1.0}$             &                   &  
 ${0.247}_{-0.002\,(0.005)}^{+0.002\,(0.004)}$  &  
 ${0.246}_{-0.002\,(0.005)}^{+0.002\,(0.004)}$  &   
 ${0.247}_{-0.002\,(0.004)}^{+0.002\,(0.004)}$  & 
 ${0.246}_{-0.002\,(0.004)}^{+0.002\,(0.004)}$  &  
 ${0.250}_{-0.004\,(0.007)}^{+0.005\,(0.008)}$  \\
\hline     
 $R_{1.4}$                  &  km             & 
 ${12.79}_{-0.28\,(0.56)}^{+0.29\,(0.55)}$    &  
 ${12.71}_{-0.30\,(0.60)}^{+0.28\,(0.54)}$    &  
 ${12.93}_{-0.27\,(0.53)}^{+0.26\,(0.48)}$    &  
 ${12.85}_{-0.27\,(0.54)}^{+0.26\,(0.48)}$    &       
 ${13.01}_{-0.34\,(0.67)}^{+0.34\,(0.60)}$    \\
 $\Lambda_{1.4}$            &             &  
 ${571}_{-87\,(162)}^{+106\,(207)}$       &  
 ${547}_{-87\,(163)}^{+101\,(203)}$       &   
 ${622}_{-89\,(164)}^{ +96\,(185)}$       &  
 ${598}_{-87\,(162)}^{ +91\,(174)}$       &  
 ${610}_{-97\,(176)}^{+109\,(212)}$       \\
 $\rho_{1.4}$               &  fm$^{-3}$        & 
 ${0.384}_{-0.036\,(0.062)}^{+0.039\,(0.078)}$  &  
 ${0.392}_{-0.038\,(0.065)}^{+0.039\,(0.080)}$  &   
 ${0.364}_{-0.028\,(0.048)}^{+0.033\,(0.068)}$  &  
 ${0.371}_{-0.029\,(0.049)}^{+0.033\,(0.069)}$  &  
 ${0.387}_{-0.037\,(0.065)}^{+0.039\,(0.075)}$  \\
 $P_{1.4}$                  & MeV/fm$^{3}$        &  
 ${51.927}_{-6.725\,(11.458)}^{+7.378\,(15.549)}$ & 
 ${53.504}_{-6.910\,(12.027)}^{+7.822\,(16.897)}$ &  
 ${48.216}_{-5.196\,( 8.919)}^{+6.288\,(13.259)}$ & 
 ${49.616}_{-5.327\,( 9.099)}^{+6.508\,(13.863)}$ &  
 ${51.114}_{-6.310\,(11.027)}^{+7.346\,(15.012)}$ \\
 $\varepsilon_{1.4}$        & MeV/fm$^{3}$        & 
 ${389.96}_{-39.23\,(66.82)}^{+42.17\,(85.26)}$   &
 ${398.48}_{-40.73\,(69.93)}^{+43.12\,(87.93)}$   &  
 ${367.88}_{-30.02\,(51.54)}^{+35.52\,(73.48)}$   & 
 ${375.12}_{-30.77\,(52.47)}^{+36.16\,(74.78)}$   &       
 ${394.15}_{-40.79\,(70.56)}^{+43.01\,(82.13)}$   \\
 $c_{s, 1.4}^2$             &                     &  
 ${0.376}_{-0.025\,(0.054)}^{+0.016\,(0.026)}$    & 
 ${0.378}_{-0.023\,(0.053)}^{+0.016\,(0.025)}$    & 
 ${0.383}_{-0.019\,(0.045)}^{+0.012\,(0.019)}$    & 
 ${0.385}_{-0.018\,(0.043)}^{+0.012\,(0.019)}$    &   
 ${0.354}_{-0.036\,(0.067)}^{+0.033\,(0.052)}$    \\
 $\Delta_{1.4}$             &                     & 
 ${0.200}_{-0.005\,(0.010)}^{+0.004\,(0.008)}$    & 
 ${0.199}_{-0.005\,(0.011)}^{+0.005\,(0.008)}$    &
 ${0.202}_{-0.004\,(0.009)}^{+0.004\,(0.007)}$    & 
 ${0.201}_{-0.004\,(0.009)}^{+0.004\,(0.007)}$    &  
 ${0.203}_{-0.006\,(0.011)}^{+0.005\,(0.009)}$    \\
\hline     
 $R_{2.0}$                  & km                &
 ${12.60}_{-0.55\,(1.14)}^{+0.51\,(0.88)}$      & 
 ${12.49}_{-0.55\,(1.14)}^{+0.51\,(0.91)}$      & 
 ${12.88}_{-0.45\,(0.94)}^{+0.39\,(0.70)}$      & 
 ${12.78}_{-0.45\,(0.95)}^{+0.39\,(0.69)}$      &    
 ${12.58}_{-0.60\,(1.20)}^{+0.54\,(0.95)}$      \\
 $\Lambda_{2.0}$            &                   &
 ${48.9}_{-16.5\,(28.6)}^{+20.1\,(38.6)}$       & 
 ${45.6}_{-15.4\,(26.9)}^{+19.4\,(37.9)}$       & 
 ${59.6}_{-16.0\,(29.6)}^{+17.5\,(33.3)}$       & 
 ${56.2}_{-15.3\,(28.1)}^{+16.8\,(31.6)}$       &    
 ${45.8}_{-17.2\,(28.3)}^{+21.8\,(42.0)}$       \\
 $\rho_{2.0}$               & fm$^{-3}$           &
 ${0.539}_{-0.078\,(0.126)}^{+0.110\,(0.254)}$    & 
 ${0.555}_{-0.082\,(0.133)}^{+0.113\,(0.260)}$    & 
 ${0.492}_{-0.054\,(0.090)}^{+0.075\,(0.177)}$    & 
 ${0.505}_{-0.057\,(0.092)}^{+0.078\,(0.182)}$    &    
 ${0.565}_{-0.095\,(0.149)}^{+0.134\,(0.292)}$    \\
 $P_{2.0}$                  & MeV/fm$^{3}$         &
 ${144.69}_{-33.64\,(51.71)}^{ +54.57\,(143.77)}$  & 
 ${152.24}_{-36.25\,(56.25)}^{ +57.61\,(152.18)}$  &  
 ${124.22}_{-22.00\,(35.18)}^{ +33.65\,(86.56)}$   & 
 ${129.65}_{-23.45\,(37.17)}^{ +35.82\,(91.55)}$   &    
 ${155.65}_{-41.29\,(62.01)}^{+70.83\,(178.64)}$  \\
 $\varepsilon_{2.0}$        & MeV/fm$^{3}$          &
 ${582.67}_{-95.26\,(150.38)}^{+140.79\,(339.60)}$  & 
 ${601.98}_{-100.50\,(160.27)}^{+144.60\,(348.88)}$ &  
 ${525.43}_{ -64.55\,(105.20)}^{ +92.48\,(225.36)}$ & 
 ${539.74}_{ -67.98\,(109.20)}^{ +96.25\,(232.28)}$ & 
 ${617.46}_{-118.56\,(182.19)}^{+178.13\,(401.64)}$ \\
 $c_{s, 2.0}^2$             &                     & 
 ${0.557}_{-0.016\,(0.036)}^{+0.019\,(0.045)}$    & 
 ${0.562}_{-0.017\,(0.035)}^{+0.021\,(0.050)}$    & 
 ${0.554}_{-0.014\,(0.030)}^{+0.015\,(0.033)}$    & 
 ${0.557}_{-0.014\,(0.029)}^{+0.015\,(0.035)}$    & 
 ${0.555}_{-0.021\,(0.045)}^{+0.023\,(0.057)}$    \\
 $\Delta_{2.0}$             &                     & 
 ${0.085}_{-0.027\,(0.065)}^{+0.020\,(0.033)}$    & 
 ${0.081}_{-0.028\,(0.068)}^{+0.022\,(0.035)}$    & 
 ${0.097}_{-0.019\,(0.045)}^{+0.015\,(0.025)}$    & 
 ${0.093}_{-0.020\,(0.047)}^{+0.015\,(0.025)}$    & 
 ${0.081}_{-0.033\,(0.077)}^{+0.023\,(0.037)}$    \\
\hline     
 $M_{\rm max}$              & $M_{\odot}$       & 
 ${2.32}_{-0.14\,(0.24)}^{+0.14\,(0.24)}$       & 
 ${2.30}_{-0.13\,(0.23)}^{+0.14\,(0.24)}$       & 
 ${2.39}_{-0.12\,(0.23)}^{+0.11\,(0.19)}$       & 
 ${2.37}_{-0.12\,(0.22)}^{+0.11\,(0.19)}$       &   
 ${2.27}_{-0.14\,(0.23)}^{+0.16\,(0.28)}$       \\
 $R_{M_{\rm max}}$          & km                &
 ${11.44}_{-0.43\,(0.78)}^{+0.49\,(0.88)}$      & 
 ${11.35}_{-0.42\,(0.76)}^{+0.48\,(0.89)}$      & 
 ${11.70}_{-0.40\,(0.77)}^{+0.40\,(0.71)}$      & 
 ${11.62}_{-0.40\,(0.75)}^{+0.39\,(0.70)}$      &   
 ${11.37}_{-0.41\,(0.70)}^{+0.50\,(0.93)}$      \\
 $\Lambda_{M_{\rm max}}$    &                   &
 ${5.64}_{-0.51\,(0.73)}^{+0.89\,(2.18)}$       & 
 ${5.67}_{-0.50\,(0.74)}^{+0.81\,(2.02)}$       & 
 ${5.34}_{-0.31\,(0.48)}^{+0.56\,(1.52)}$       & 
 ${5.36}_{-0.31\,(0.48)}^{+0.55\,(1.43)}$       &   
 ${6.04}_{-0.81\,(1.14)}^{+1.24\,(2.67)}$       \\
 $\rho_{M_{\rm max}}$       & fm$^{-3}$         & 
 ${0.912}_{-0.084\,(0.142)}^{+0.091\,(0.169)}$  & 
 ${0.928}_{-0.086\,(0.147)}^{+0.089\,(0.166)}$  & 
 ${0.864}_{-0.063\,(0.108)}^{+0.074\,(0.152)}$  & 
 ${0.877}_{-0.065\,(0.109)}^{+0.075\,(0.151)}$  &  
 ${0.935}_{-0.096\,(0.162)}^{+0.094\,(0.162)}$  \\
 $P_{M_{\rm max}}$          & MeV/fm$^{3}$      &
 ${551.63}_{-34.96\,(63.29)}^{+34.92\,(69.28)}$ & 
 ${559.77}_{-35.57\,(66.13)}^{+35.43\,(72.67)}$ &
 ${534.23}_{-30.03\,(53.29)}^{+32.30\,(64.15)}$ & 
 ${541.30}_{-30.83\,(54.22)}^{+32.07\,(65.21)}$ & 
 ${554.04}_{-35.56\,(65.38)}^{+36.80\,(71.19)}$ \\
 $\varepsilon_{M_{\rm max}}$  &  MeV/fm$^{3}$        & 
 ${1189.94}_{-111.16\,(188.08)}^{+116.94\,(219.07)}$ & 
 ${1209.51}_{-112.67\,(194.16)}^{+114.86\,(216.09)}$ & 
 ${1126.41}_{ -83.56\,(142.85)}^{ +97.23\,(198.06)}$ & 
 ${1143.10}_{ -85.54\,(144.32)}^{ +98.06\,(196.78)}$ & 
 ${1222.63}_{-128.27\,(216.29)}^{+123.06\,(214.00)}$ \\
 $c_{s, M_{\rm max}}^2$     &                     & 
 ${0.741}_{-0.028\,(0.063)}^{+0.018\,(0.028)}$    & 
 ${0.739}_{-0.025\,(0.058)}^{+0.018\,(0.028)}$    & 
 ${0.751}_{-0.019\,(0.048)}^{+0.012\,(0.019)}$    & 
 ${0.750}_{-0.019\,(0.045)}^{+0.012\,(0.019)}$    &  
 ${0.732}_{-0.034\,(0.068)}^{+0.026\,(0.038)}$    \\
 $\Delta_{M_{\rm max}}$     &                     & 
 ${-0.134}_{-0.015\,(0.023)}^{+0.022\,(0.049)}$   & 
 ${-0.133}_{-0.015\,(0.023)}^{+0.020\,(0.046)}$   & 
 ${-0.143}_{-0.010\,(0.015)}^{+0.016\,(0.039)}$   & 
 ${-0.142}_{-0.010\,(0.015)}^{+0.016\,(0.036)}$   &
 ${-0.125}_{-0.022\,(0.032)}^{+0.027\,(0.053)}$   \\  
\hline                                                                                  
 $R_{1.4}$-$R_{2.0}$        & km             & 
 ${0.18}_{-0.23\,(0.37)}^{+0.34\,(0.75)}$    & 
 ${0.20}_{-0.24\,(0.39)}^{+0.33\,(0.74)}$    &
 ${0.04}_{-0.16\,(0.27)}^{+0.22\,(0.53)}$    & 
 ${0.06}_{-0.16\,(0.27)}^{+0.23\,(0.54)}$    & 
 ${0.40}_{-0.38\,(0.61)}^{+0.52\,(1.07)}$   \\
 $R_{1.4}$-$R_{1.0}$        & km             & 
 ${0.12}_{-0.13\,(0.24)}^{+0.12\,(0.20)}$    & 
 ${0.11}_{-0.12\,(0.23)}^{+0.12\,(0.21)}$    & 
 ${0.18}_{-0.10\,(0.21)}^{+0.09\,(0.16)}$    & 
 ${0.17}_{-0.10\,(0.21)}^{+0.09\,(0.16)}$    & 
${-0.03}_{-0.23\,(0.41)}^{+0.23\,(0.38)}$    \\
\hline\hline
\end{tabular}
\end{table*}
\endgroup

In this appendix, we present the characteristic parameters of nuclear
matter at saturation density and key gross quantities of CSs 
predicted by CDFs under different astrophysical scenarios. 
The results obtained from Scenarios B are listed in Tables~\ref{tab:NM_Posterior_B}
and~\ref{tab:CS_Posterior_B}, and those from Scenarios F in Tables~\ref{tab:NM_Posterior_F}
and~\ref{tab:CS_Posterior_F}.

To ensure the presentation is self-contained, we give here the 
definitions for nuclear matter characteristic coefficients. 
For the isoscalar sector, an order $n$ coefficient can be defined as 
\begin{align}
X^n_{\rm sat} = 
3^n \rho^n_{\rm sat}\frac{\partial^n \varepsilon\,(\rho, 0)}{\partial\, \rho^n}\Bigg\vert_{\rho_{\rm sat}},
\end{align}
where $\varepsilon\,(\rho, 0)$ is the energy density of symmetric matter.
It gives at saturation density the energy per nucleon $E_{\rm sat}$ $(n = 0)$, 
the incompressibility $K_{\rm sat}$ $(n = 2)$, 
the skewness $Q_{\rm sat}$ $(n = 3)$, 
and the kurtosis $Z_{\rm sat}$ $(n = 4)$, respectively. 
For the the isovector sector, the symmetry energy is computed from
\begin{align}
E_{\rm sym}(\rho) = 
\frac{1}{2}\frac{\partial^n \varepsilon\,(\rho, \delta)}{\partial\, \delta^2}\Bigg\vert_{\delta = 0}.
\end{align}
Similarly, it can be expanded close to the saturation density and an 
order $n$ coefficient can be written as 
\begin{align}
X^n_{\rm sym} = 
3^n \rho^n_{\rm sat}\frac{\partial^n E_{\rm sym}(\rho)}{\partial\,\rho^n}\Bigg\vert_{\rho_{\rm sat}},
\end{align}
with $J_{\rm sym}$ $(n = 0)$ the symmetry energy coefficient at 
saturation density, the slope $L_{\rm sym}$ $(n = 1)$, the 
curvature $K_{\rm sym}$ $(n = 2)$, the skewness $Q_{\rm sym}$ $(n = 3)$, 
and the kurtosis $Z_{\rm sym}$ $(n = 4)$, respectively. 

Note that in Tables~\ref{tab:NM_Posterior_B} and~\ref{tab:NM_Posterior_F} 
we also show the values of higher order parameters $Z_{\rm sat}$ in the
isoscalar sector and $K_{\rm sym}$, $Q_{\rm sym}$ and $Z_{\rm sym}$ in
the isovector sector, which are predictions based on values of the
lower-order parameters that fix the parameters of the present CDF.


%
\end{document}